\documentclass[twocolumn]{emulateapj}

\usepackage{amsmath}

\shortauthors{
Wong et al.
}

\shorttitle{
Gas Flow within the Bondi Radius of NGC~3115
}
\slugcomment{Astrophysical Journal, submitted}

\begin{document}
\title{
The Megasecond {\it Chandra} XVP Observation of NGC~3115: 
Witnessing the Flow of Hot Gas within the Bondi Radius
}

\author{
Ka-Wah Wong\altaffilmark{1},
Jimmy A. Irwin\altaffilmark{1},
Roman V. Shcherbakov\altaffilmark{2,3},
Mihoko Yukita\altaffilmark{1,4},
Evan T. Million\altaffilmark{1},
and
Joel N. Bregman\altaffilmark{5}
}

\altaffiltext{1}{Department of Physics and Astronomy, University of 
Alabama, Box 870324, Tuscaloosa, AL 35487, USA
}
\altaffiltext{2}{Department of Astronomy, University of Maryland, 
College Park, MD 20742-2421, USA
}
\altaffiltext{3}{Hubble Fellow
}
\altaffiltext{4}{Johns Hopkins University, Homewood Campus, Baltimore, 
MD, 21218, USA
}
\altaffiltext{5}{Department of Astronomy, University of Michigan, 500 
Church Street, Ann Arbor, MI 48109-1042, USA
}

\email{kw6k@email.virginia.edu}

\begin{abstract}

Observational confirmation of hot accretion model predictions has been 
hindered by the challenge to resolve spatially the Bondi radii of black 
holes with X-ray telescopes.  Here, we use the Megasecond {\it Chandra} 
X-ray Visionary Project (XVP) observation of the NGC~3115 supermassive 
black hole to place the first direct observational constraints on the 
spatially and spectroscopically resolved structures of the X-ray 
emitting gas inside the Bondi radius of a black hole.  We measured 
temperature and density profiles of the hot gas from a fraction out to 
tens of the Bondi radius ($R_B =$ 2\farcs4--4\farcs8 = 112--224~pc).  
The projected temperature jumps significantly from $\sim0.3$~keV beyond 
5\arcsec\ to $\sim 0.7$~keV within $\sim 4\arcsec$--5\arcsec, but then 
abruptly drops back to $\sim 0.3$~keV within $\sim 3$\arcsec.  This is 
contrary to the expectation that the temperature should rise toward the 
center for a radiatively inefficient accretion flow.  A hotter thermal 
component of $\sim 1$~keV inside 3\arcsec\ ($\sim 150$~pc) is revealed 
using a two component thermal model, with the cooler $\sim$0.3~keV 
thermal component dominating the spectra.  We argue that the softer 
emission comes from diffuse gas physically located within $\sim 150$~pc 
from the black hole.  The density profile is broadly consistent with 
$\rho \propto r^{-1}$ within the Bondi radius for either the single 
temperature or the two-temperature model.  The X-ray data alone with 
physical reasoning argue against the absence of a black hole, supporting 
that we are witnessing the onset of the gravitational influence of the 
supermassive black hole.

\end{abstract}


\section{Introduction}
\label{sec:intro}

Understanding how supermassive black holes accrete matter from their 
galactic surroundings is an important, yet still poorly understood 
process. While spectacular in nature, quasars accreting at $\sim$$10\%$ 
of their Eddington limit with luminosities of $\sim$$10^{46}$ ergs 
s$^{-1}$ do not represent the current behavior of the vast majority of 
supermassive black holes. Even more mildly accreting ($\sim$$10^{-5} 
L_{\rm Edd}$) black holes classified as AGN only constitute a few 
percent or less of the supermassive black hole population depending on 
environment \citep[e.g.,][]{DG83,HB92,Ho08,Ho09}. Instead, nearly all 
supermassive black holes exhibit a much more modest ($<10^{-8} L_{\rm 
Edd}$) radiatively inefficient accretion mode, notably illustrated by 
the quiescent 4 million solar mass black hole at the center of the Milky 
Way.

The well known classical Bondi accretion model \citep{Bon52} suggests 
that in order to be accreted, gas must be within a distance from the 
black hole where the gravitational potential of the black hole dominates 
the thermal energy of the hot gas. The ``sphere of influence" for gas 
around a black hole is defined by its Bondi radius, $R_B = 2GM_{\rm 
BH}/c_s^2$, where $M_{\rm BH}$ is the mass of the black hole, and $c_s$ 
is the sound speed of gas far away from the black hole. For a billion 
solar mass black hole with hot gas temperature of $\sim 0.1$--1~keV, the 
Bondi radius is on the order of tens to hundreds of parsecs, or 5--6 
orders of magnitude greater than the Schwarzchild radius of the black 
hole ($R_S = 2GM_{\rm BH}/c^2$). Although realistic astrophysical 
accretion may be dramatically different from the Bondi accretion model 
due to its simple idealized assumptions, studying hot gas properties 
within the Bondi ``sphere of influence" remains crucial for 
understanding how matter is being accreted.

It is not the case that very low-luminosity black holes are simply 
starved for gas. For example, the Bondi rate of gas flowing through the 
Bondi radius of Sgr A* at the center of the Milky Way \citep[$\dot M$$_B 
\sim 10^{-6}$ M$_{\odot}$ yr$^{-1}$;][]{Bag+03} would imply a luminosity 
of $\sim10^{41}$~ergs~s$^{-1}$ at the standard 10\% radiative efficiency 
\citep[e.g.,][]{FR95}, several orders of magnitude higher than is 
observed \citep[][and references therein]{NMG+98}. Two general solutions 
have been proposed to account for the missing radiative energy. One 
solution is that although matter passing through the Bondi radius makes 
it to the event horizon of the black hole, most of the energy in the gas 
is carried by the ions, and is advected down the black hole before 
radiating much energy \citep[advection-dominated accretion flows, or 
ADAFs;][]{Ich77,RBB+82,NY94}. The second solution is that matter passing 
through the Bondi region does not make it to the event horizon of the 
black hole, but either circulates in convective eddies 
\citep[convective-dominated advection flows, or 
CDAFs;][]{NIA00,QG00,AIQ+02}, or some of the gas actually escapes the 
potential of the black hole in an outflow \citep[such as 
advection-dominated inflow-outflow solutions, or ADIOS;][]{BB99,Beg12}, 
or variations on these themes.

Ideally, one would like to compare the predictions of radiatively 
inefficient accretion flow models with the X-ray--determined properties 
of the hot gas flowing from the galactic potential into the Bondi region 
\citep{BM99,QN00}. Most notably the 
temperature and density profiles of the hot gas provide leverage for 
distinguishing among competing accretion flow models. However, 
observational confirmation of predictions of these theories has been 
hindered by the inability to resolve spatially the Bondi radii of black 
holes with X-ray telescopes. For even the closest, most massive black 
holes, the angular size of the Bondi regions are on the order of only a 
few arcseconds or less \citep{Gar+10}. Sgr A* is by far the 
best-studied Bondi region both observationally and theoretically 
\citep[e.g.,][]{YMF02,YQN03,Bag+03,SB10,Wan+13}, but 
with a detected size of only $1\farcs5$ in X-ray,
simultaneous spatial and 
spectral analysis 
is challenging. Few Bondi regions with radii exceeding 
2$^{\prime\prime}$ are accessible with {\it Chandra}, and these 
candidates suffer from the presence of a bright point source in or near 
the nucleus of the galaxy (M87) or low X-ray gas count rates (NGC~3115), 
or both (M31*).
 
Despite its low X-ray count rate, the gas surrounding the black hole in 
NGC~3115 provides us with the best opportunity to obtain 
spatially-resolved spectral information on the hot gas within the Bondi 
region of a black hole. At a distance of 9.7~Mpc \citep{TDB+01}, 
NGC~3115 is the nearest galaxy with a one to two billion solar mass 
black hole \citep{Kor+96,EDB99}. The low temperature ($\sim$0.3~keV) of the 
ambient gas implies a Bondi radius of $R_B = 112$--224~pc = 
$2\farcs4$--$4\farcs8$ \citep[][hereafter W11]{WIY+11}. Previous moderate length {\it 
Chandra} observations of NGC~3115 revealed evidence for an increase in 
the hot gas temperature inside the Bondi region of its supermassive 
black hole (W11), one of the tell-tale signatures of most 
(non-cooling) accretion flow models. W11 also found the slope 
of the density of the hot gas inside the Bondi radius to be $\rho \sim 
r^{-1}$, although neither the temperature spike nor the density slope 
could be constrained to high significance owing to the low X-ray count 
rate. The tantalizing results prompted a deeper 1 Megasecond {\it 
Chandra} X-ray Visionary Project (XVP) observation of NGC~3115 to 
collect the required number of X-ray photons to derive temperature and 
density profiles on a spatial scale that matches the resolution of {\it 
Chandra}.
 
Here we describe the results from our analysis of the {\it Chandra} 
Megasecond observation of NGC~3115. After careful subtraction of 
contaminating X-ray emission from other sources, we derive the first 
spatially-resolved temperature and density profiles of gas inside the 
Bondi region of a black hole. In a companion paper \citep{SWI+13}, we 
develop radial gas flow models for NGC~3115.

The paper is organized as follows.  Section~\ref{sec:obs} describes the 
X-ray observations and data analysis.  In Section~\ref{sec:AGN}, we 
examine the spatially extended nature of the hot gas within the Bondi 
radius and address the lack of evidence of X-ray emission 
from the central weak AGN.  
Section~\ref{sec:results} describes the observational results,
in particular, the temperature, surface brightness, and density profiles 
of the hot gas, as well as an unexpected strong soft emission within the 
Bondi radius, and evidence of a (at least) two-temperature structure of 
the hot gas within $\sim 150$~pc from the black hole.
We discuss possible origins of the central soft emission 
in Section~\ref{sec:soft}.  
We argue against the idea that spun-up stars can be an important 
X-ray component at the NGC~3115 center 
in Section~\ref{sec:spunupstars}.  The implications to 
accretion models are discussed in Section~\ref{sec:models}.  We 
summarize and conclude in Section~\ref{sec:summary}.  Systematic 
uncertainties in spectral modeling is addressed in detail in 
Appendix~\ref{sec:app1}.  The X-ray upper limit of the central weak AGN
is assessed in Appendix~\ref{sec:appAGN}.

At a distance of 9.7~Mpc, the angular scale of NGC~3115 is 47~pc/1\arcsec.
Unlike our previous paper (W11) which presents $1\sigma$ 
confidence, errors are given at 90\% confidence level in this paper 
unless otherwise specified.

\section{X-ray Observations and Spectral Analysis}
\label{sec:obs}

\subsection{Data Reduction}
\label{sec:reduction}

\begin{figure*}
\includegraphics[width=1.\textwidth, angle=0]{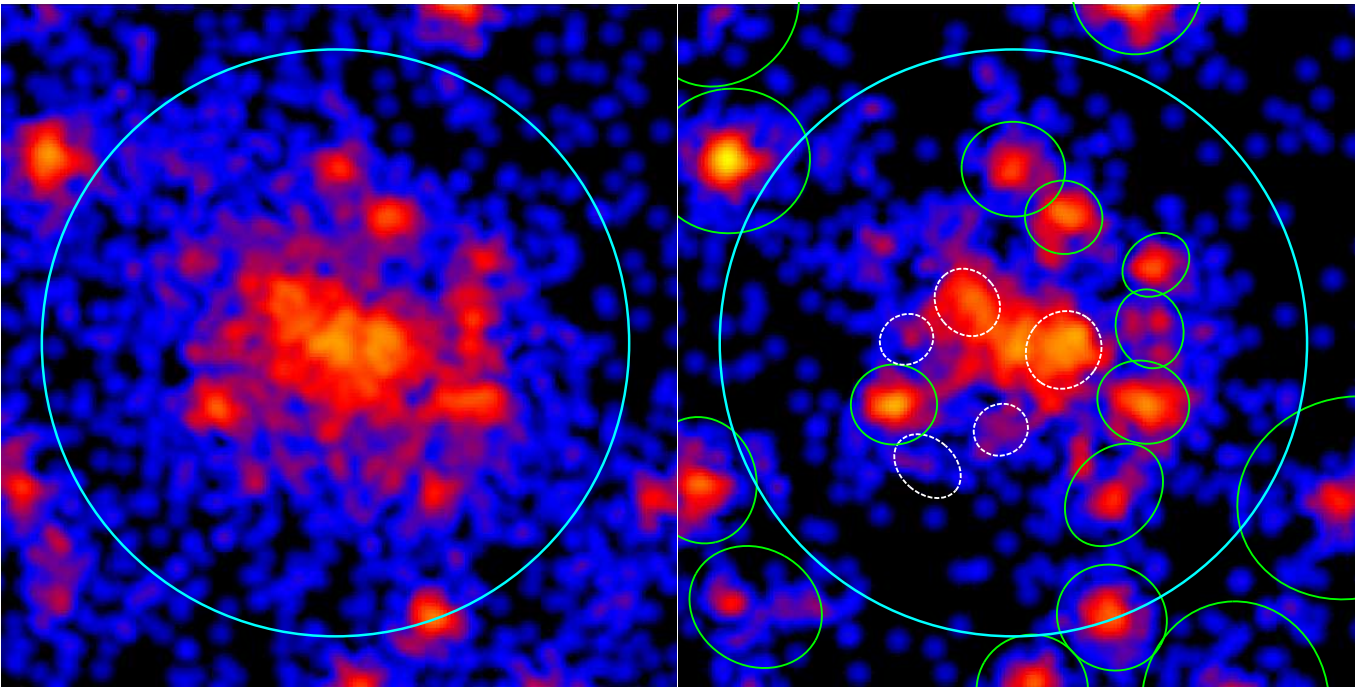}
\caption{
Smoothed soft band (0.5--1.0~keV: left) and hard band (2.0--6.0~keV: 
right) {\it Chandra} sub-pixel resolution images of NGC~3115 with 1 
image pixel binning size = 0\farcs0615. Both images were smoothed with 
a Gaussian kernel of FWHM = 0\farcs3.  North is top while east is 
left.  The large 
circles (cyan) are centered on the extended emission; each has a 
radius of 5\arcsec (=235~pc).  The point sources (or compact structures) 
removed are shown in smaller solid green and dashed white ellipses, 
with the dashed white ellipse 
sources only detected on sub-pixel resolution images (see text).  The 
unresolved diffuse emission in the soft band image is dominated by hot 
gas beyond $\gtrsim 2\arcsec$ while the hard band image is dominated by 
LMXBs.  Note the more extended and rounder extended emission in the soft 
band compared to the narrower extended structure of the hard emission 
inside 2\arcsec--3\arcsec, with the orientation of the hard emission 
roughly aligned with the major axis of the optical emission along the NE-SW 
direction \citep{Kor+96}.
}
\label{fig:image}
\end{figure*}

NGC~3115 (Figure~\ref{fig:image}) was observed eight times with the {\it 
Chandra} in 2012 between January and April (ObsIDs 13817, 13819, 13820, 
13821, 13822, 14383, 14384, and 14419) for a total of 998~ks.  Unless 
otherwise specified, we only included all these 2012 observations in the 
data analysis but do not include the 155~ks observations taken in 2001 
(ObsID 2040) and 2010 (ObsIDs 11268 and 12095) because the effective 
area has changed dramatically even since 2010 due to the increasing 
Advanced CCD Imaging Spectrometer (ACIS) 
contamination.  We note that including the 2001 and 2010 observations 
generally introduced a $\lesssim$10--20\% systematic error in the spectral analysis 
but does 
not improve the statistical uncertainties.  
In all the observations NGC~3115 was placed near the ACIS-S aimpoint.
All the data were 
reprocessed using the {\tt chandra\_repro} script of {\tt CIAO 4.4} and 
{\tt CALDB 4.4.10}.  
The default sub-pixel event-repositioning algorithm ``EDSER" 
was used.  After removed flares using the {\tt CIAO} {\tt deflare} script, the 
cleaned exposure time was 972~ks.

To improve the astrometry between different observations required for 
the high spatial resolution analysis, we have performed relative 
astrometry correction for each observation.  We first created a 
sub-pixel resolution image in 0.3--6.0 keV with a binning size of 0.123\arcsec\
(0.25 ccd pixel) for a $\sim 4\times 4$ square arcmin region 
around NGC~3115 for each observation.  We then used the {\tt CIAO} {\tt 
wavdetect} script to create an initial source list for each image.  
This source list was only used for astrometry correction.  The 
longest observation (ObsID 13820) was used as the reference for the 
relative astrometry corrections.  The {\tt CIAO} {\tt reproject\_aspect} script 
was then used to create new aspect solutions for all the other 
observations.  All of the data were reprocessed again using {\tt 
chandra\_repro} with the new aspect solutions to complete the astrometry 
corrections.

We extracted a local background from a 70\arcsec--90\arcsec\ annular 
region far enough from the center of NGC~3115 so that the source-removed surface 
brightness of the X-ray emission is basically flat.  
Background contributes negligibly to the inner $\sim 10\arcsec$ 
but becomes significant only in the outermost regions 
(Appendix~\ref{sec:app1}).  
Changing the background level by $\pm 
10\%$ introduces systematic uncertainties that are much smaller than the 
statistical uncertainties within 20\arcsec.  It only introduces a 
systematic uncertainty that is larger than the statistical uncertainty in 
gas temperature beyond 20\arcsec, although the gas normalization is still 
hardly affected.

\subsection{Point Source Removal}
\label{sec:ptsrc}

To analyze the diffuse X-ray emission of the hot gas it is necessary to remove
contaminating point sources (or compact 
structures).  Point sources were detected with {\tt CIAO 
wavdetect}.  To detect as many point sources as possible, we used all 
the observations except the data taken in 2001 because its optical axis 
position is significantly different ($> 1$\arcmin) from the other 
observations.  We created images with 1 ccd pixel binning size in four 
energy bands (0.3--1.0, 1.0--2.0, 2.0--6.0, and 0.3--6.0 keV) and 
combined images according to energy bands in these observations.  
The source regions in different energy 
bands were then visually inspected and combined.  We refined the region 
sizes of the point sources (or compact
structures) detected within 4\arcsec\ by using sub-pixel 
images with 0.125 pixel binning size and ran {\tt CIAO wavdetect} again.  
With these sub-pixel images, a few more weak sources and also some elongated 
structures were identified within 3\arcsec\ (Figure~\ref{fig:image}).  
Unless otherwise specified, we have removed all these structures except 
for the source detected at the galaxy center in our nominal data 
analysis.
Including or removing these structures 
within 3\arcsec\ 
gives essentially the same results for the gas component.

As mentioned above, the central peak was detected with {\tt wavdetect}.  
Our analysis shows no strong evidence of a point source and the central peak
is clearly extended (Section~\ref{sec:AGN}).  
We have determined the upper limit of the potential point source and
found that hot gas measurements are not affected by this potential weak AGN
(Section~\ref{sec:AGN} and Appendix~\ref{sec:appAGN}).  
Therefore, we did not remove the central region and 
we ignored any potential AGN contamination in our analysis.

\subsection{Spectral Analysis}
\label{sec:spectra}

We extracted spectra in circular annuli centered on the central peak of 
the extended X-ray emission (0.3--6.0~keV),
which is assumed to be the center of the flow (the supermassive black 
hole).  This peak is within $\lesssim 0\farcs05$ of the soft 
(0.3--2.0~keV) emission peak.  It is separated by 0\farcs15 from the 
optical peak we measured using the archival {\it HST} data (below), 
consistent with the position uncertainty.
The diffuse gas 
distribution is assumed to be spherically symmetric which is justified in 
Section~\ref{sec:SB}.  
All the spectra were analyzed using the X-ray Spectral Fitting 
Package\footnote{http://heasarc.nasa.gov/xanadu/xspec/} ({\tt XSPEC}).

The unresolved X-ray emission is mainly contributed by unresolved 
low-mass X-ray binaries (LMXBs), stellar emission from 
cataclysmic variables and coronally active binaries (CV/ABs), and the 
diffuse hot gas component that we are interested in. After the CV/AB component
is spectrally-subtracted in a statistical manner (described below), the very soft
emission from the gas and the
very hard emission from the unresolved LMXBs can be reliably separated 
through spectral fitting.
The combined spectra of resolved low-$L_X$ 
($<10^{37}$ ergs s$^{-1}$) LMXBs in the bulge of M31 are very similar to 
more luminous LMXBs \citep{IAB03}. Therefore, we can assume the unresolved LMXB 
emission to be spectrally modeled as the brighter resolved sources. 
With our deep Megasecond observation, many more LMXBs were detected than in W11, and 
the unresolved LMXBs are no longer the dominant component at 
$\gtrsim 2\arcsec$ in the 0.5--1.0~keV band (Figure~\ref{fig:surbri} below).
We modeled the LMXB component as a power law model and fixed the 
power-index to $\Gamma_{\rm LMXB} = 1.6$ which is consistent with the 
value of $1.61^{+0.02}_{-0.02}$ measured from the combined spectrum of 
all the resolved point sources within $D_{25}$ of NGC~3115 and the value 
(1.6) of the summed emission from many resolved X-ray binaries in nearby 
galaxies \citep{IAB03}.
Using $\Gamma_{\rm LMXB} = 1.4$ or 1.8 gives essentially the same 
results for the analysis of the hot gas (Appendix~\ref{sec:app1}).

The faint and soft sources similar to those of the Galactic Ridge 
emission (CV/AB) contribute appreciately to the X-ray flux beyond the 
Bondi region.  
Hence, including this component is essential in the 
analysis.
In W11, we used the 2MASS K-band image to estimate the CV/AB 
contribution, and assumed the X-ray flux of the CV/AB component scales
linearly with the K-band luminosity.  Because the typical spatial resolution of 2MASS is about 
2\arcsec--3\arcsec\, which is poorer than the {\it Chandra} resolution, 
the CV/AB in the central regions can be underestimated.

In this paper, we use a higher resolution {\it HST} WFPC2 I-band (F814W 
filter) image to estimate the
CV/AB contribution.  We assume the intrinsic K-band surface brightness 
profile follows the I-band surface brightness profile.  The archival {\it HST} 
I-band image has a PSF FWHM and a pixel size of $\sim$0\farcs1 
which is much smaller than 
the 2MASS PSF (2\arcsec--3\arcsec) and is close to the {\it Chandra} 
PSF ($\sim$0\farcs5 near the aimpoint).  Therefore, we can 
use the {\it HST} I-band surface brightness profile to estimate 
the intrinsic K-band surface brightness profile.  In practice, we first 
smoothed the {\it HST} I-band image with a Gaussian kernel.  We then re-scaled 
the {\it HST} I-band image to match the 2MASS flux unit within a radius of 
10\arcsec\ centered at the surface brightness peak.  After that,
we generated a surface brightness profile with a radial binning 
size of 1\arcsec\ within 10\arcsec\ for the {\it HST} I-band image and 
compared to the 2MASS surface brightness profile.  We found that with a 
Gaussian kernel of 2\farcs5 FWHM, the smoothed {\it HST} I-band surface 
brightness profile matches the 2MASS profile moderately well.  The 
deviations between the two profiles are at most 17\% at all radii which are 
smaller then the conservative 50\% uncertainty of the CV/AB contribution 
we considered in this paper (see below).  
The unsmoothed {\it HST} I-band image was then 
scaled with the same factor as the smoothed {\it HST} image and was used to 
estimate the intrinsic K-band surface brightness.  With this correction, 
the flux within the central 1\arcsec\ is higher than the 2MASS estimate 
by about a factor of two.

To model the X-ray spectrum of the CV/AB component, we fitted the unresolved X-ray emission of the
dwarf elliptical galaxy M32, 
which is believed to be hot gas-free \citep{RCS+07, BKF11}.
Using an absorbed thermal + power 
law [{\tt PHABS*(APEC+POWERLAW)}] model fitted to  
archival {\it Chandra} data,
the best-fit temperature was $T_{\rm 
CV/AB}=0.76^{+0.07}_{-0.06}$ keV and the power law index was $\Gamma_{\rm 
CV/AB} = 1.92^{+0.07}_{-0.11}$.
The 
CV/AB normalizations of each annulus were determined by the $L_X$--$L_K$ 
scaling relation derived from M32.
We investigated the systematic uncertainties by varying the CV/AB 
normalizations by $\pm 50\%$, comparable to the galaxy by galaxy 
variation of this component (Appendix~\ref{sec:app1}).  This only changed the measured gas 
properties slightly within $\sim 10\arcsec$.  All of the systematic 
uncertainties are smaller than the statistical uncertainties.

\begin{figure*}
\includegraphics[width=2.5truein, angle=270]{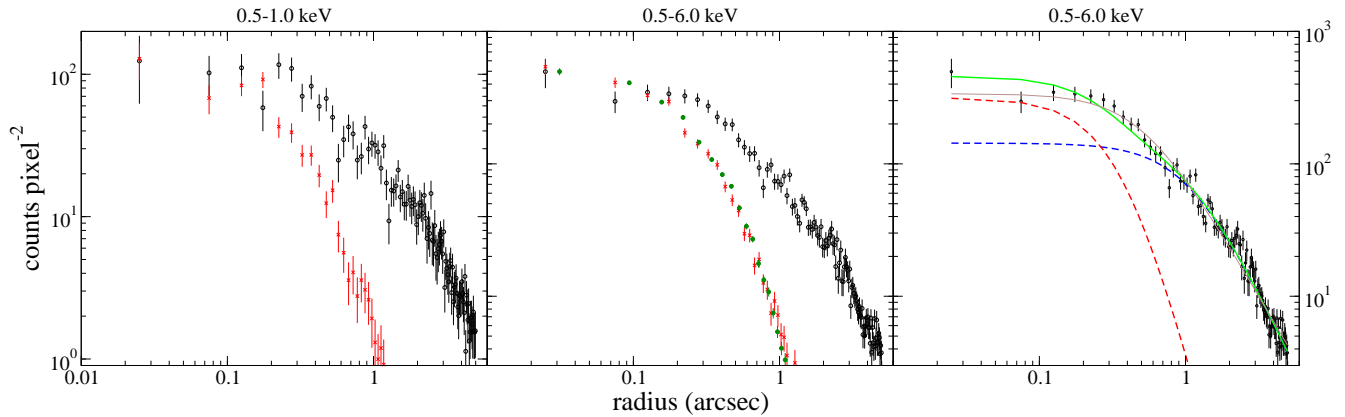}
\caption{
Left panel: Surface brightness profiles of the unresolved diffuse 
emission (black open circles) 
and a nearby point source $6\arcsec$ away from the 
center (red crosses) in 0.5--1.0~keV.  
The point source profile was normalized to the 
photon counts of the unresolved diffuse profile within 0\farcs2.
This point source is used as the PSF template. 
Middle panel: Similar as the left panel but in 0.5--6.0~keV.  
The green solid circle data are from 
another very bright point source about 30\arcsec\ from the galaxy 
center.
Right panel: Surface brightness profiles of the unresolved diffuse 
emission (black open circles).  
The sum of the two-component (Moffat+Moffat) model is 
shown in thick solid green, with the individual components of the point source 
(red dashed) and extended source (blue dashed) shown in dashed lines 
(See Appendix~\ref{sec:appAGN} below).  The single 
Moffat model is shown as a thin solid brown line.
All the error bars in this figure are at the $1\sigma$ confidence level.
}
\label{fig:psfcompare}
\end{figure*}

The eight observations in 2012 were observed in three different periods 
(two observations between January 18--January 23, three observations between 
January 26--February 5, and three observations between April 4--April 7).  
For each period, the pointings and roll angles of the observations are 
nearly identical.  Therefore, we merged the spectra for observations 
taken in each period using the {\tt CIAO specextract} script.  The three 
merged spectra of each extraction region in the 0.5--6.0~keV energy 
range were fitted jointly to the three component absorbed ({\tt PHABS}) 
model -- a thermal ({\tt APEC}) model for the gas, a power law ({\tt 
POWERLAW}) with a slope of 1.6 for the unresolved LMXBs, and a combination of thermal + 
power law ({\tt APEC + POWERLAW}) model for the CV/ABs.  We fixed the 
absorption at the Galactic value of $N_{\rm H} = 4.32 \times 
10^{20}$~cm$^{-2}$ \citep{DL90}.  The systematic uncertainty introduced 
by $N_{\rm H}$ is not significant (Appendix~\ref{sec:app1}). 
We fitted the temperature of the thermal gas 
component.  The normalizations of the LMXB component were allowed to 
vary in the three merged spectra to account for possible 
variability.  For the thermal gas component, when the three 
normalizations are untied in the joint fitting, they are within the 
uncertainties of each other.  Moreover, the ccd responses did 
not change much during the observations and the extended hot gas should 
not be time varying.  Therefore, we tied the hot gas normalizations in 
the fitting. 
The metallicity was fixed to the solar value using the {\tt wilm} 
abundance table \citep{WAM00} and
the systematic uncertainties of this assumption are discussed in Appendix~\ref{sec:app1}. 
In brief, thawing the metallicity only introduces a small systematic bias in hot 
gas temperature compared to the statistical uncertainty and increases 
the gas normalizations by a factor of four to five,
and the derived gas density only increases by a 
factor of two without affecting the density slope (Appendix~\ref{sec:app1}).  None of 
our conclusions are sensitive to the adopted metallicity.
Unless otherwise specified, all the spectra were fitted 
using the c-statistic.

\section{Spatially Extended Emission in the Bondi Region and  
X-ray Limits of the Weak AGN}
\label{sec:AGN}

Recent radio observations have detected a point source at the center of 
NGC~3115 with a luminosity of $L_{\rm 8.5 GHz} = 3.1 \times 
10^{35}$~erg~s$^{-1}$, suggesting nuclear activity of the 
supermassive black hole \citep{WN12}.  It is important to constrain the 
luminosity of a potential weak AGN in X-ray, either in understanding the 
radiative process around the vicinity of the black hole or the potential 
contamination to extended emission of larger scale thermal hot gas. 
Earlier studies suggested a point-like source at the center of NGC~3115 
and measured an X-ray luminosity up to about $4\times 
10^{38}$~erg~s$^{-1}$ \citep[e.g.,][]{Ho09,BKF11}.  Such a high 
luminosity would explain all the X-ray emission within the central 
1\arcsec\ region (although it would hardly be able to contaminate the 
Bondi region between $\sim1$\arcsec\ and 5\arcsec\ in radius).  However, 
W11 and \citet{MGT+12} argued that any point source emission 
should be significantly weaker due to the blending of extended emission. 
W11 provided an upper limit of $10^{38}$~ erg~s$^{-1}$ for any central point source.  In 
this section, we quantify the extended emission within the Bondi region 
and provide stricter (while still conservative) limits on the central AGN.

Figure~\ref{fig:psfcompare} shows the surface brightness profile of the 
central 5\arcsec\ region and also the normalized profile of a nearby 
point source.  Using other nearby sources gives consistent point source 
profiles.  This figure strongly suggests that the X-ray emission is 
extended beyond a fraction of an arcsec, with no strong evidence of a 
central point source.  
The X-ray emission within the central 1\arcsec\ in radius varied 
$\lesssim 2\sigma$ in four different energy bands (0.5--1.0, 1.0--2.0, 
2.0--6.0, and 0.5--6.0~keV) in all the eleven observations, showing no 
evidence of a varying central point source.
Spectral analysis suggests that hot gas 
contributes about half of the soft emission (0.5--1.0~keV) at 
1--2\arcsec\ and is the dominant component beyond that 
(Figure~\ref{fig:surbri} below), and therefore the left panel of 
Figure~\ref{fig:psfcompare} indicates a clearly extended hot gas component 
beyond sub-arcsec scale.

By modeling the spatial distribution of the X-ray emission within 5\arcsec\ 
with a two-component model (a point source component and an extended 
diffuse component; Appendix~\ref{sec:appAGN}), we found that the 
conservative upper limit of the X-ray luminosity of the AGN is $L_{X, 
{\rm AGN}} < 4.4\, (1.1) \times 10^{37}$~erg~s$^{-1}$ in 0.5--6.0 
(0.5--1.0)~keV. This is about two to nine times lower than the 
quoted detection 
or upper limits determined recently 
(e.g., \citealt{Ho09,BKF11}; W11; \citealt{MGT+12}).  We determined the upper 
limit of the Eddington fraction to be $L_{X, \rm AGN}/L_{\rm Edd} < 3.5 
\times 10^{-10} (10^9 M_{\sun}/M_{\rm BH})$, making it one the most 
under-luminous AGNs \citep{Ho08}.  
Therefore, the accretion of the NGC~3115 black
hole is expected to be in the hot mode with an expected temperature
profile close to the virial temperature of the system and increasing
toward the center. 
We also found that the AGN at most 
contributes $\sim 30\%$ to the X-ray emission 
within a radius of 1\arcsec\ in 0.5--6.0~keV.  Such a systematic 
uncertainty will not affect our results of the hot gas profiles 
qualitatively (Appendix~\ref{sec:appAGN}). Therefore, we ignore the AGN 
contribution in our data analysis.

\section{Spatially Resolved Hot Gas Properties}
\label{sec:results}

\subsection{Temperature Profile}
\label{sec:tprofile}

\subsubsection{Single temperature model}
\label{sec:1T}

\begin{figure}
\includegraphics[width=3.5in, angle=0]{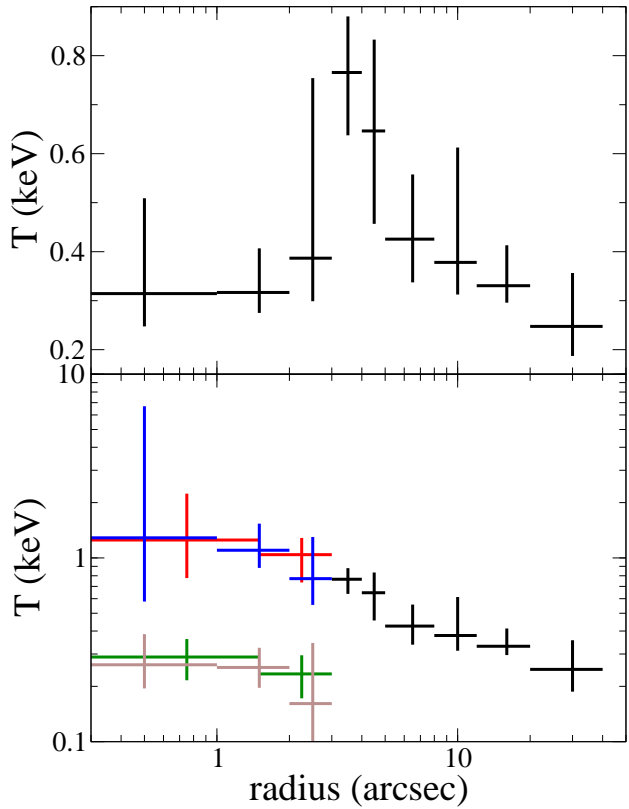}
\caption{
Upper panel: Temperature profile using a single temperature model. Lower 
panel: Temperature profile using a two-component model within 3\arcsec\,
with color data points representing each temperature component.
The corresponding temperatures of the hotter (red/blue) and cooler (green/brown) components of 
the two-temperature model represent two different radial binning schemes.
There is no evidence of two-(or 
multi-)temperature structure beyond 3\arcsec, and the black data points beyond 3\arcsec\
are the same single temperature shown in the upper panel. For both 
panels, vertical error bars are at the 90\% confidence level and horizontal 
bars indicate the radial binning size.
}
\label{fig:t_profile}
\end{figure}

We model the projected spectra of the diffuse gas component with a single 
temperature optically thin thermal plasma model ({\tt APEC}) described 
in Section~\ref{sec:spectra}.  The projected temperature profile is 
shown in the upper panel of Figure~\ref{fig:t_profile}.  It is
clear that the projected temperature is around 0.3~keV in the outer 
region and increases sharply to about 0.7~keV within about 5\arcsec. 
Within the inner 2\arcsec\ or 3\arcsec, the new data now 
strongly suggest that the projected temperature drops at the center, 
indicating that there is significant soft emission near the center.  
Previous analysis of the moderate length 2001+2010 {\it Chandra} data by W11 suggested an increase in best-fit 
projected temperature toward the center, although the uncertainty was too 
large to be conclusive.  Such a simple interpretation of a monotonic 
increase in projected temperature is no longer valid.  We also noticed 
that a single temperature fit to the projected temperature within the 
inner $\sim 3\arcsec$ is no longer adequate.  In Section~\ref{sec:2T} 
below, we present two-temperature fitting results to the data.

\subsubsection{Two-temperature model}
\label{sec:2T}

\begin{figure}
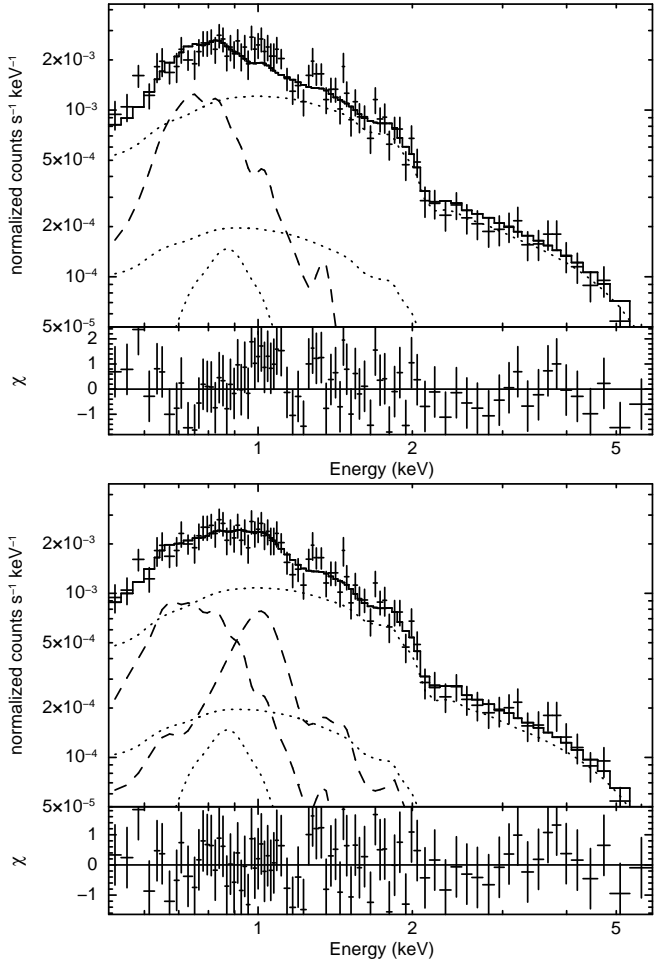

\includegraphics[width=2.5in, angle=270]{./f4a.ps}
\includegraphics[width=2.5in, angle=270]{./f4b.ps}
\caption{
Upper panel: Single temperature fit to the spectrum in the 1\arcsec--3\arcsec\ 
annular region.  The dashed line is the thermal ({\tt APEC}) component.  
The top dotted line is the LMXB component.  The two lower dotted lines 
are the CV/AB component.  
The solid line is the sum of all the components.
A clear residual can be seen at around 1~keV. 
Lower panel: Similar to the upper panel but with the extra thermal ({\tt 
APEC}) component of the two-temperature fit labeled with a second dashed 
line.
All the error bars in this figure are at $1\sigma$ confidence level.
}
\label{fig:spectra}
\end{figure}

Single temperature fitting to the projected spectra within the inner 
3\arcsec\ suggests a central drop in temperature.  However, a single 
temperature model may not be sufficient to characterize the projected 
spectra; at the very least the spectra should consist of gas at different 
temperatures due to projection effects of gas at larger radii.  
Motivated by the expected central rise in 
temperature for radiatively inefficient accretion flows (RIAFs) 
accreting in hot modes 
\citep[e.g.,][]{NY94,FR95,BM99,NM08,GM13} and evidence of central 
temperature peaks $\lesssim 300$~pc from the galactic nuclei in a few 
early-type galaxies \citep{PBF+03,HBB+08,PWF+12}, we searched for 
possible evidence of a hot thermal component potentially hidden in a 
multi-temperature structure within the central 3\arcsec.  We first 
examined the spectra with different binning sizes 
including or removing structures within 3\arcsec.  We 
also combined all the different observations taken in 2012 to create a 
single spectrum in each extraction region, so that the combined spectra 
have enough counts for visual inspection.  This also allows us to group 
the spectra with a minimum of 25 counts per spectral bin to use 
chi-squared or F-statistics.  We found that while a single temperature 
model generally gives a good enough fit ``globally'' ($\chi^2_{\nu} 
\approx 1$) for most spectra, there is a notable systematic excess of 
emission at about 1~keV.  An example of a spectrum in an annular region 
of 1\arcsec--3\arcsec\ is shown in Figure~\ref{fig:spectra}.  The 
best-fit temperature of a single temperature model is 
$0.37^{+0.11}_{-0.06}$~keV with $\chi^2=75.9$ and 80 degree of freedom.  
When we added one more thermal component (an {\tt APEC} model with the same 
abundance and redshift as the first thermal component), this 
two-temperature model gave best-fit temperatures of 
$1.23^{+0.25}_{-0.21}$ and $0.29^{+0.05}_{-0.05}$~keV with $\chi^2=53.6$ 
and 78 degree of freedom. The lower panel of Figure~\ref{fig:t_profile} 
shows the spectrum of the two-temperature model.  A simple F-test with 
the F-statistic of 16.3 and probability of $1.26\times 10^{-6}$ strongly 
suggests that the second component is needed.  A more formal test for an 
additional component was also performed by simulating 1000 spectra with 
the single temperature model and then comparing the likelihood ratio of 
the single temperature model with respect to the two-temperature model 
(likelihood ratio test in {\tt XSPEC}).  We found that all 1000 of the simulated 
likelihood ratios are smaller than the observed ratio, strongly suggesting 
that the two-temperature model is preferred.  We performed a similar 
likelihood ratio test to determine whether the extra component is a narrow 
single line emission or a broader thermal component by simulating 1000 
spectra with an extra {\tt gaussian} model in {\tt XSPEC}.  The line width 
was fixed to zero.  We found that 99.2\% of the simulated likelihood 
ratios of the single temperature + {\tt gaussian} model with respect to 
the two-temperature model are smaller than the observed ratio, again, 
strongly favoring the two-temperature model.  We conclude that there is 
evidence of a hotter thermal component with temperature 
$\gtrsim 1$~keV within the central 3\arcsec\ region.

It is possible that there is a wider distribution of temperature 
structure along the line of slight.  Unfortunately, the statistics of 
our data do not allow us to test beyond a two-temperature model.  It 
is also known that a multi-temperature structure is difficult to be 
quantified from the data \citep[e.g.,][]{KPD+08,Gay13}.  Here, we simply 
characterize the thermal component within 3\arcsec\ with a 
two-temperature model.  This two-temperature model at least may be able 
to characterize the rough lower and upper limits of the temperature 
distribution.  Temperature profiles of the two-temperature model are 
shown in the lower panel of Figure~\ref{fig:t_profile}.  The fittings 
were performed by joint-fitting different observations as described in 
Section~\ref{sec:spectra} rather than fitting the combined spectra.  
Different colors represent different spatial binning.  It is interesting 
that the hotter temperature rises all the way toward the center, 
consistent with most hot accretion models.  
Considering the hotter component within 3\arcsec\ and the single 
temperature profile beyond that, fitting the projected temperature 
profile to a power law gives $T \propto r^{-[0.44^{+0.29}_{-0.33}]}$ 
(90\% confidence) for $r < 5\arcsec$ and $T \propto 
r^{-[0.34^{+0.25}_{-0.25}]}$ for $5\arcsec < r < 40\arcsec$.  Ignoring 
the central 1\arcsec\ region gives essentially the same results.
It is also interesting that 
the lower temperature of the two-temperature model is consistent with the 
low temperature of $\sim 0.3$~keV outside the Bondi radius.

Since it is quite certain that there are at least two (or multiple) 
temperature structures within about 3\arcsec, we also tested whether there 
is any evidence of significant two-temperature structure beyond 
3\arcsec. 
Between 3\arcsec\ and 5\arcsec, the best-fit 
temperatures and normalizations of the hotter component are consistent 
with a single temperature fit; the best-fit lower temperature is 
consistent with zero, which is unphysically low and 
the flux (0.5--2.0~keV) of the soft thermal 
component is less than 10\% of the hot component.  Beyond 5\arcsec, 
either the higher and lower temperatures are consistent with each other 
within 90\% confidence, or the best-fit temperatures are more sensitive 
to systematic uncertainties.  The best-fit temperatures ($\sim 0.3$~keV) and 
normalizations of the cooler component are consistent with single 
temperature fits.  The cooler component dominates over the hot component 
by a factor of about two to four in normalization.
Performing additional tests by tying the higher (and lower) temperature 
of a few radial bins together does not change the conclusion.  We conclude that 
a single temperature model is sufficient to approximate the thermal 
structure of the hot gas beyond 3\arcsec.

\subsection{Surface Brightness Profiles}
\label{sec:SB}

\begin{figure}
\includegraphics[width=2.8in, angle=270]{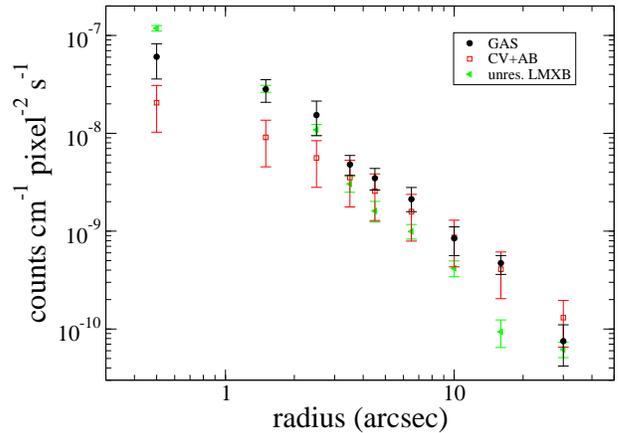}
\caption{
Surface brightness profiles for the hot gas (black circles), CV/AB (red 
squares), and LMXB (green triangles) components in the 0.5--1.0~keV 
band for NGC~3115.  One pixel equals 0\farcs492.
The error bars for the hot gas and LMXB are at the 90\% confidence level.  
The error bars for the CV/AB component are the conservative 
uncertainties of $\pm 50\%$ we assumed in addressing the systematic 
uncertainties.
}
\label{fig:surbri}
\end{figure}

The surface brightness profiles for the hot gas, the CV/AB, and the 
unresolved LMXB components are shown in Figure~\ref{fig:surbri}.  Unlike 
the surface brightness profiles in W11 which assume all the 
emission above 2~keV is contributed by the LMXBs, in this paper, the 
surface brightness of the hot gas and unresolved LMXB were calculated 
from the best-fit models spectroscopically.  This self-consistently 
takes into account the uncertainties in the hot gas and LMXB 
contributions.  We use a single temperature model for the hot gas in 
this plot.  
In the central region where the gas is multi-temperature, 
this single temperature model, which has a low temperature 
of $\sim 0.3$~keV near the center (Section~\ref{sec:1T}), should 
be able to take 
into account most of the gas emission of the hot gas in the 0.5--1~keV 
band.  Using a two-temperature model gives essentially the same result,
but gives larger uncertainties.

The hot gas is robustly 
detected within $\sim$20\arcsec--40\arcsec.  Hot gas is the dominant component 
in this energy band out to $\sim 10\arcsec$--20\arcsec.  The CV/AB 
component is not significant within a few arcsec but becomes significant 
in the outermost regions.  However, varying the CV/AB contributions by 
$\pm 50\%$ does not change any of our conclusions qualitatively at all 
radii and does not significantly change any result quantitatively within 
a few arcsec (Appendix~\ref{sec:app1}).

The distribution of optical light of NGC~3115 is highly elliptical.  We 
have tested whether the thermal X-ray emission deviates from azimuthal 
symmetry by extracting surface brightness profiles of the hot gas in 
four 90-degree sectors in NW, NE, SE, and SW directions.  We conclude 
that there is no evidence of azimuthally variation for the single 
temperature model and the hot component of the two-temperature model, 
and therefore spherical approximation is adequate for our analysis.  
However, there is some weak evidence that the cooler component of the 
two-temperature model is distributed more along the major axis of the 
galaxy within 3\arcsec\ and this implication is discussed in 
Section~\ref{sec:smalldisk}.  Nevertheless, we also present systematic 
tests by assuming the gas is distributed elliptically as the optical 
light and as a thick disk in Section~\ref{sec:soft:projection}.

\subsection{Emission Measure and Density Profiles}
\label{sec:den}

\begin{figure}
\includegraphics[width=3.5in, angle=0]{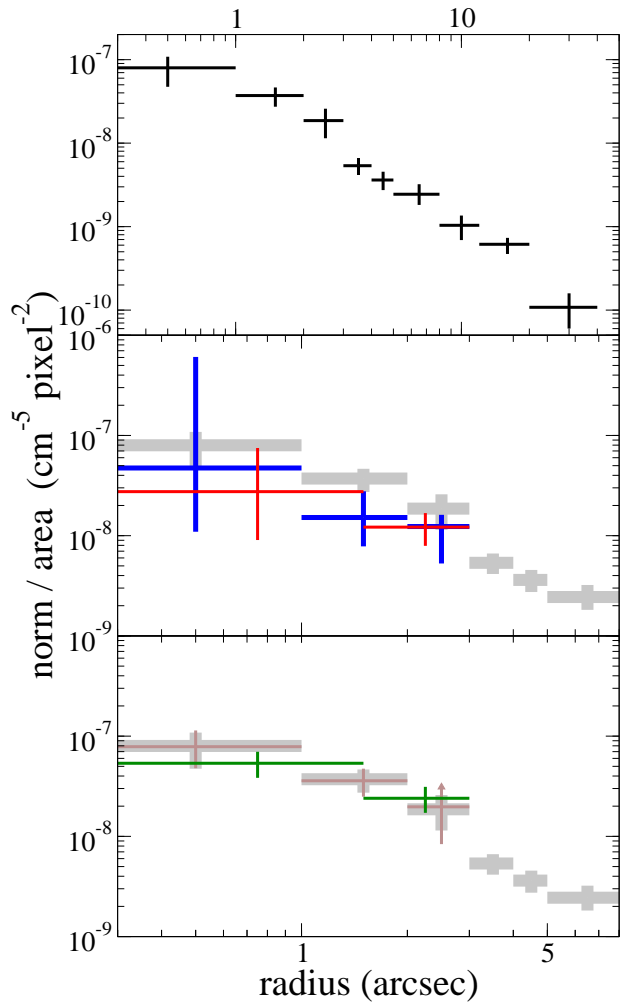}
\caption{
Upper panel: {\tt APEC} normalization per unit area of the single 
temperature model within 40\arcsec.  One pixel equals  
0\farcs492. Middle panel: {\tt APEC} normalization per unit surface area of 
the hot component of the two-temperature model within 8\arcsec\ are shown 
in color data points, with different colors for different radial 
binnings.  The thick grey data points are for the same single temperature 
model shown in the upper panel.  
Lower panel: Similar to 
the middle panel but with the low temperature component instead of the 
hot component. 
For all panels, vertical error bars are at the 90\% 
confidence level and horizontal bars indicate the radial binning size,
with the exception that 
the upper limit of the data point of the cold component (lower panel) 
at 2\farcs5 is not well determined 
and is labeled as a triangle.
}
\label{fig:surnorm}
\end{figure}

Figure~\ref{fig:surnorm} shows the {\tt XSPEC} {\tt APEC} normalization per 
unit surface area, which is proportional to the emission measure of 
$\int n_e^2 dl$, where $n_e$ is the electron density and $l$ is the 
column length along the line of sight.  The single temperature model is 
shown in black in the upper panel and thick grey in the middle and 
lower panels.  The hotter (cooler) component of the two-temperature 
model within 3\arcsec\ is shown in color in the middle (lower) panel, 
with different colors representing different spatial binning.  
In general, the emission measure of the hotter component is lower than the 
single temperature model but barely consistent within the error bars 
while the cooler component is comparable to the single temperature 
model.
This indicates that the hotter gas density may be slightly lower and 
the cooler gas density may be similar to that determined by a single 
temperature model (as shown below), but the density profiles should not 
be too sensitive to these two models (since $n_e \propto \sqrt{\rm emission\,\, 
measure}$). In the central 1\farcs5, the uncertainties of the hot 
component appear to be significantly larger than 
that of the single temperature model.  The large error bar is due to the poor 
temperature constraint in that region, with a higher temperature upper 
limit so that the gas normalization is degenerate with the hard 
emission from, e.g., LMXBs.  We also noted that there is a weak AGN in 
the central 1\arcsec\ which can contribute at most up to 
30\% of the total emission and can increase the hot 
component uncertainties.  
Note that this upper limit is very conservative (see
Appendix~\ref{sec:appAGN}).

Within 3\arcsec, the soft normalization per unit area decreases 
with radius (lower panel in Figure~\ref{fig:surnorm}), suggesting that the 
the cooler component should not be projected emission from a much larger 
relatively-uniform background structure.  The centrally peaked soft 
normalization per unit area suggests that the cooler component 
should be located physically inside about 150~pc (3\arcsec) from the 
galaxy center (see Section~\ref{sec:soft} for more detailed discussions).

With the {\tt XSPEC} {\tt APEC} normalizations (or emission measure) in each 
annulus, we can deproject the density profile using the onion peeling 
method as was done in W11 and also outlined in detail in 
\citet{KCC83} or \citet{WSB+08}.  In brief, this technique calculates 
the emission measure of each spherical shell starting from the outermost 
annulus toward the center, and the emission measure of each 
subsequent shell is calculated by subtracting the projected emission 
measure from the outer shells.  Unlike W11 who deprojected 
the density from the surface brightness and assumed a certain spectral shape 
(or gas temperature), in this work, we directly deprojected the density 
profile from the spatially resolved emission measure fitted from 
spectra, and therefore the uncertainties of the spectral shape (or 
temperature) have been partially taken into account.
Note that the deprojected density determined here is effectively the 
root-mean-squared of the density. If the filling factor is less than one 
or if the gas is not homogeneous, our deprojected density is 
overestimated \citep[see also][]{SWI+13}.
Unfortunately, X-ray observation alone cannot determine the filling 
factor or clumpiness.  Theoretical models may provide constraints to 
these factors.

\begin{figure}
\includegraphics[width=3.5in, angle=0]{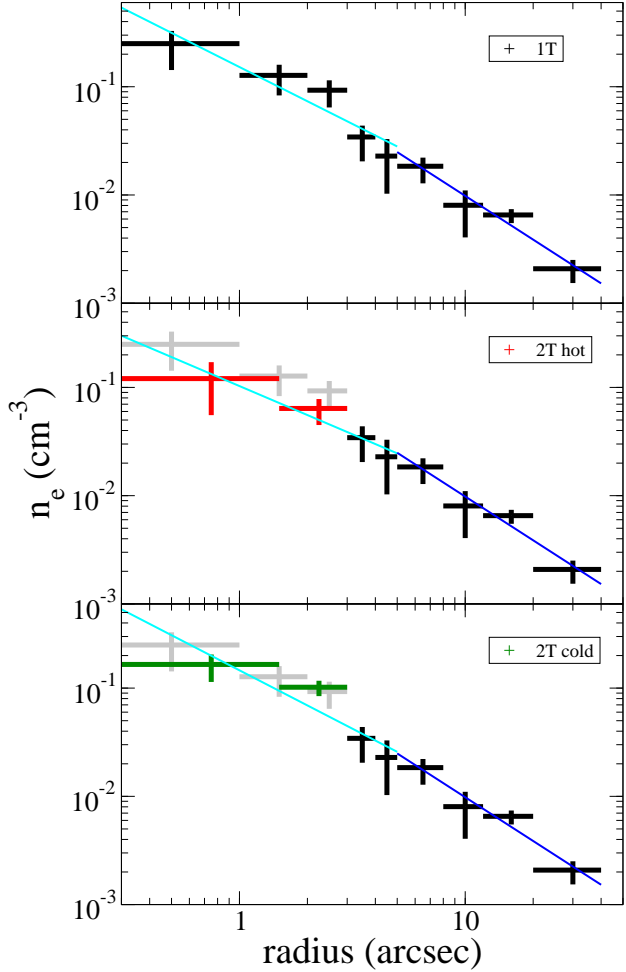}
\caption{
Upper panel: Deprojected density profile of the single temperature model 
(black). The cyan (blue) line has a power law index of 1.05 (1.34) in 
0\arcsec--5\arcsec\ (5\arcsec--40\arcsec). 
Middle panel: Deprojected density profile using the 
hot component of the two-temperature model within 3\arcsec\ (red).  
Single temperature model (black) were used beyond 3\arcsec.  The density 
profile of the single temperature mode within 3\arcsec\ is shown in grey 
for comparison. The cyan (blue) line has a power law index of 0.89 
(1.34) in 0\arcsec--5\arcsec\ (5\arcsec--40\arcsec). 
Lower panel: Similar to the middle panel, 
but with the hot component replaced by the cold component (green) of the 
two-temperature model. 
The cyan (blue) line has a power law index of 
1.08 (1.34) in 0\arcsec--5\arcsec\ (5\arcsec--40\arcsec). 
For all panels, vertical error bars 
are at the 90\% confidence level and horizontal bars indicate the radial 
binning size.
}
\label{fig:nprofile}
\end{figure}

The deprojected electron density profiles of the single and 
two-temperature models are shown in Figure~\ref{fig:nprofile}.  The 
errors were estimated by running $10^6$ Monte Carlo simulations.  The 
density profile of the single temperature model (upper panel of 
Figure~\ref{fig:nprofile}) is easier to interpret by assuming a 
single-phase plasma.  However, the two-temperature model is more 
difficult to interpret.  It is possible that the two components are in 
two (or more) distinct phases or the particles can be distributed in a 
broader than Maxwellian distribution.  In any case, the distribution of 
the two-temperature model cannot be constrained without additional 
assumptions (e.g., filling factor, gas distribution).  
Motivated by hints that the colder component may be 
distributed as a ($\sim 3$\arcsec) disk along the major axis of the 
galaxy and also by the fact that the hotter component appears to be more spherical,
(see Sections~\ref{sec:SB} and \ref{sec:smalldisk}), 
we assume a simple model that 
the hotter component characterizes the spherically distributed hot gas 
in projection within 3\arcsec\ (i.e., assuming a filling factor
of 1 for the hotter component).  We assume the cold component is 
concentrated in a small disk-like region that can be ignored when doing 
the spherical deprojection analysis.  The origin of such a cold 
component is discussed in detail in Section~\ref{sec:soft}.  With the 
two-temperature model within 3\arcsec, the emission measure 
(normalization) cannot be constrained well enough if we thaw the 
temperatures.  In particular, using a narrow spatial binning size of 1\arcsec\ 
gives too large statistical uncertainties and also the systematic 
uncertainties in the central 1\arcsec\ can be larger. To improve the 
constraints, we used a larger spatial bin of 1.5\arcsec\ for deprojection. 
We fixed all the 
higher and lower temperatures to their best-fit values and assessed the 
uncertainties of the normalizations.  
The deprojected density of the 
hotter component is shown in the middle panel of 
Figure~\ref{fig:nprofile}.  As expected, it is slightly lower than the 
single temperature model.  We noted that if we assume the cold component 
as the spherically distributed gas and ignore the hot component (lower 
panel of Figure~\ref{fig:nprofile}), the density profile is closer 
to the single temperature model. Note that the density profiles under 
these three different assumptions are remarkably similar, suggesting 
that these models are measuring similar emission measure that is more 
sensitive to density than temperature.
Note also that if the filling factor is less than one or if the gas is 
clumpy, the density we measured is biased high.

Fitting the density profile of the hotter component of the 
two-temperature model within $5\arcsec$ to a power law gives $\rho 
\propto r^{-[0.89^{+0.35}_{-0.45}]}$ (90\% confidence; note that 
W11 present $1\sigma$ confidence).
The less physically motivated density profile of the cooler component of 
the two-temperature models gives $\rho \propto 
r^{-[1.08^{+0.31}_{-0.24}]}$  in 0\arcsec--5\arcsec.
The single temperature model gives 
a power law index of
$1.05^{+0.25}_{-0.25}$ in 0\arcsec--5\arcsec, 
$0.90^{+0.24}_{-0.30}$ in 0\arcsec--4\arcsec, 
and
$0.62^{+0.26}_{-0.38}$ in 0\arcsec--3\arcsec.  
The density profile becomes steeper in 
the $5\arcsec$--$40\arcsec$ outer region, with a power law index of 
$1.34^{+0.20}_{-0.25}$.

\section{Origin of the Soft Emission within 150~pc}
\label{sec:soft}

The single temperature model suggests that there is significant soft 
emission with a characteristic temperature of $\sim 0.3$~keV within 
$\sim 150$~pc (3\arcsec).  
The preference for a two-temperature model strongly suggests that 
there is at least a two (or multiple) temperature structure.  We 
discuss the origin of this soft emission in the following sections.
We argue below that it is unlikely that the soft emission emanates from 
projected gas from larger radii, or from soft stellar sources, but 
instead is most likely explained by cooler thermal gas located 
physically within $\sim$150~pc.

\subsection{Insufficient Projected Cooler Gas from the Outer Region}
\label{sec:soft:projection}

Because the characteristic temperature ($\sim 0.3$~keV) of the softer 
emission is very similar to the temperature of the outer region beyond 
$\sim 5$\arcsec, 
it is possible that the softer emission comes from
projected gas of the outer regions.  However, the surface 
brightness (or emission measure per unit area) is quite steep in the 
central 3\arcsec\ (lower panel in Figure~\ref{fig:surnorm}), which does 
not appear to come from projected gas of a larger outer region.  It 
would be ideal if we could fit a projection model (such as the {\tt XSPEC} 
{\tt projct}, although it has some limitation in the assumed geometry) 
to test whether projection can account for all the soft emission. 
Unfortunately, the statistics of the data do not allow us to perform 
such a test with high confidence.  Instead, we performed three 
conservative tests to quantify the allowed projected soft emission.

We first tested whether the soft emission within 3\arcsec\ can be 
projected from a spherical distribution of $\sim 0.3$~keV gas beyond 
5\arcsec.  To maximize the projection effect, we assumed that all gas 
beyond 5\arcsec\ has a temperature of 0.3~keV.  This may overestimate 
the cool gas contribution as the hot gas temperature between 
$\sim$5\arcsec--10\arcsec\ is slightly hotter and most of the projection 
should come from this region.  We then fitted the gas normalizations in 
each annulus beyond 5\arcsec\ with the gas temperature fixed to 0.3~keV.  
By using the onion peeling method (Section~\ref{sec:den}; \citet{KCC83}; 
\citet{WSB+08}), the projected gas (flux) contribution to the inner 
3\arcsec\ region can be calculated.  In the 1\arcsec--3\arcsec\ region, 
we fitted a two-temperature model with the lower temperature fixed to 
0.3~keV to estimate the total flux of the cooler component.  The inner 
1\arcsec\ is ignored due to the potential contamination from a weak AGN.  
We found that projected gas can only account for 11\% of the soft 
emission in the 1\arcsec--3\arcsec\ annulus.  Even if the soft emission 
is at its lower limit of the 90\% confidence interval (95\% one-sided 
limit), projected gas can only account for 16\% of the soft emission.  
The statistical uncertainty of the projected gas normalization is of the 
order of 20--30\%, and therefore the uncertainty in projected gas cannot 
account for the difference.  

We then constructed an oblate spheroid model of the gas halo with 
constant ellipticity which roughly follows the optical light 
\citep[minor radius/major radius = 0.6;][]{KR92}
Here, we extracted spectra in elliptical 
annular regions with the radial binning sizes along the major axis equal 
to our circular annulus sizes.  By doing a similar analysis as the 
spherical model above, we found that projected gas can still at most 
account for about 22\% of the soft emission 
in the 1\arcsec--3\arcsec\ annulus.  

Finally, we assume a thick circular disk of uniform gas with thickness 
of 6\arcsec\ and an outer radius of 40\arcsec\ aligned along the optical 
major axis.  The rotation axis is assumed to be parallel to the plane of 
the sky along the optical minor axis.  We extracted a spectrum from a $6 
\times 80$ arcsec rectangular region aligned along the major 
axis, with a 3\arcsec\ circular region at the center and point sources 
excluded.  Again, we fitted the {\tt APEC} normalization of a 0.3~keV 
gas in this region.  The projected gas within the 3\arcsec\ region is 
proportional to the projected volume.  We found that projected gas can 
only account for 9\%--13\% of the soft 
emission within the central 1\arcsec--3\arcsec\ annulus.  This disk model 
accounts for a smaller amount of projected gas compared to the ellipsoid 
model.  This may be due to the very bright optical disk in the disk region 
that overestimates the CV/AB contribution in X-ray and hence 
underestimates the projected gas.  Another reason may be that the gas 
emission is not disk-like as assumed.

In summary, most of the soft emission within the central 
3\arcsec\ cannot be explained by projected emission from a spherical, an 
oblate spheroid, or a thick disk distribution of cooler gas.  Projection 
may work, e.g., if the outer cooler gas is preferentially 
distributed toward the line-of-sight of the supermassive black hole, 
which seems unlikely.

\subsection{Difficulties of Stripped Cores of Giant Stars}
\label{sec:giantcores}

It has been suggested 
that tidally-stripped cores of giant stars can have very soft spectra ($E 
\lesssim$~keV) with relatively large luminosities ($>100 L_{\odot}$) which can 
last for $10^3$--$10^6$~yr around supermassive black holes 
\citep[e.g.,][]{DGM+01, DK05}
and could conceivably account for the soft emission inside 3\arcsec.
The tidal radius of a billion solar mass black hole is $R_t \approx 1.2 
R_S (M_{*}/M_{\odot})^{-1/3} (M_{\rm BH}/10^9 M_{\odot})^{-2/3} 
(R_{*}/5R_{\odot}) $, where $R_S$ is the Schwarzchild radius, and $M_{*}$ 
and $R_{*}$ are the stellar mass and radius, respectively \citep{MGR12}.  
Therefore, most main-sequence stars with $R_{*} \lesssim 5 R_{\odot}$ 
are directly swallowed by the billion solar mass black hole, but the 
envelopes of giant stars can be tidally stripped.  

First, for the nearest two point-like sources or extended regions located 
near the major axis at 1\arcsec, spectral analysis suggests that their 
spectra are perfectly consistent with a typical LMXB spectrum with a 
power-law index of $1.65\pm0.13$.  These weak sources do not affect the 
spectral analysis of the hot gas and do not provide excess soft emission 
in the central 3\arcsec\ region.  We have removed all the detected point 
sources within the 1\arcsec--3\arcsec\ annular region and there is no 
indication of any other point source. If the soft emission comes from 
these soft stripped cores, the luminosity of each source has to be lower 
than $\sim 10^{36}$~erg~s$^{-1}$ to remain undetected.

Second, the rate of all stars passing through the corresponding tidal 
radius (including those swallowed and disrupted) of NGC~3115 has been 
estimated to be about ${\dot N} \sim 5 \times 10^{-5}$~yr$^{-1}$ 
\citep{WM04}, which is roughly consistent with other estimations of 
giant galaxies with supermassive black holes \citep[e.g.,][]{MT99,SU99}.  
The fraction of giant stars ($R_{*} > 5$--$10R_{\odot}$) passing through 
the tidal radius is about 5--10\% \citep[e.g.,][]{MGR12}.  Taking the 
upper limit of 10\% gives the tidal stripping rate of giant stars to be 
${\dot N_G} \sim 5 \times 10^{-6}$~yr$^{-1}$.  Even if a stripped core can 
maintain its luminosity for as long as $10^6$~yr, there are only about five 
stripped cores 
luminous expected at one time around the center of NGC~3115.
The 
total unabsorbed luminosity of the soft component of the two-temperature 
model in the 1\arcsec--3\arcsec\ annular region is $L_{\rm 0.5-2~keV}=2\pm 
0.4\times 10^{37}$~erg~s$^{-1}$,
implying that at least 20 low-luminosity stripped cores are required to 
account for the soft emission. This is a factor of $>$4 greater than the 
expected number of stripped cores at any one time.

Third, from the spatial scale of $\sim 100$~pc of the soft emission and 
a typical velocity dispersion of $250$~km~s$^{-1}$, these cores would 
have traveled for 0.4~Myr.
If most of the soft emission comes from these stripped cores, the 
average lifetime of the emission has to be close to the upper limit of 
the expectation.

In summary, to account for all the soft X-rays with stripped giant cores, 
each core needs to be less luminous than $\sim 10^{36}$~erg~s$^{-1}$, 
the tidal stripping rate of giant stars needs to be higher than the 
predicted value of a few $\times 10^{-6}$~yr$^{-1}$, and the average 
lifetime of the emission should be longer than a fraction of a Myr.  
These set very tight and challenging constraints on the properties of 
the stripped cores to meet in order to explain most of the soft emission 
seen at the galaxy center.

\subsection{Multi-temperature Gas?}
\label{sec:multiT}

Given the difficulties of the projected gas and stripped cores 
scenarios, it is likely that most  
of the soft emission is physically located within the Bondi 
region.  In fact, it is expected that the interstellar medium (ISM) can be in a 
multi-temperature phase\footnote{It 
is also known that 
hot gas in other systems such as galaxy groups or clusters can be 
multi-temperature \citep[e.g.,][]{BLB+03,Tre+12}.}  
\citep[e.g.,][]{MO77,Mun+04,RSI06}.  
Simulations 
have shown that hot gas accreted toward supermassive black holes can be 
chaotic and has a wide range of temperature within the Bondi radius 
\citep{BPN12,GRO13,DS13}.  Numerical simulations also suggest that 
thermal instabilities of non-rotating cooling gas occurs when $t_{\rm 
cool}/t_{\rm ff} \lesssim 10$, where $ t_{\rm cool}$ and $t_{\rm ff}$ 
are the cooling time and free fall time scales \citep{SMQ+11,GRO13}.  
For NGC~3115, $t_{\rm cool}/t_{\rm ff} \approx 100$ \citep{SWI+13},  
so cooling may not be important to induce thermal instability.  
However, it may be that there are some regions in NGC~3115 where $t_{\rm 
cool}/t_{\rm ff} \lesssim 10$ locally and therefore cooling can become 
important.  This cooling gas may also be cooling out of the X-ray 
emitting hot phase ($T\gtrsim0.1$~keV) and therefore only the hotter 
phase with longer $t_{\rm cool}$ is detected \citep[see Figure~5c 
in][]{GRO13}.  It may also be that the gas is not free falling 
(accretion rate greatly suppressed by, e.g., rotation) and therefore the 
relevant time scale is the accretion time rather than the free fall time.  
The accretion rate of any cooling gas out of the X-ray band should, 
however, not be too high to trigger a powerful AGN at the moment.
From a theoretical point of view, such a multi-temperature phase, 
however, is less likely to be clumpy as clumpiness or fragmentation in 
accretion flow is more likely to occur with higher accretion rate 
\citep[$\eta \dot M c^2/L_{\rm Edd} \gtrsim 0.02$, where $\eta$ is the 
radiative efficiency;][]{WCL12} or larger $L_{X,{\rm AGN}}/L_{\rm Edd} 
\sim 0.01$, where $L_{X,{\rm AGN}}$ is the AGN luminosity \citep{BPN12}.

\subsection{Cooler Gas Resides in a Small Disk?}
\label{sec:smalldisk}

Hot gas in an early-type galaxy with significant stellar rotational 
velocity is likely to be rotating to some degree, in particular, if a 
significant faction of the hot gas comes from stellar mass loss.  For 
NGC~3115, the total hot gas within $10R_B$ is about $5 \times 10^6 
M_{\odot}$.  Assuming the specific stellar mass lost rate of $1.5 \times 
10^{-12}$~yr$^{-1}$ estimated by \citet{Mat89} and a total stellar mass 
of $5 \times 10^{10} M_{\odot}$ within $10\, R_B$ \citep{KR92} implies 
that only 70 Myr is needed to build up the hot gas from stellar mass 
loss, which is much shorter than the expected age of an early-type 
galaxy (on the order of 10 Gyr).  Therefore, we expect the angular momentum 
of the hot gas in NGC~3115 to be comparable to the stellar component.  In 
fact, X-ray observations and numerical simulations suggest that hot 
gas can be rotating collectively with some rapidly rotating early-type 
galaxies \citep[e.g., NGC~4649;][]{BMH+09}.  The implications of rotation 
of hot gas are discussed briefly in Section~\ref{sec:models} below.

As mentioned in Section~\ref{sec:SB}, 
there is some weak 
evidence that the cooler gas component of the two-temperature model 
is preferentially located on the major axis while the hot component is more 
spherically symmetric within the 3\arcsec\ region.  Interestingly, there 
is also a very small and distinct optical thin disk with a radius of 
about 3\arcsec\ along the major axis of the galaxy as shown with {\it 
HST} (Figure~4 in \citealt{Kor+96} and Figure~A8 in \citealt{LSD+10}).
It is possible that some of 
the gas can circularize and have enough time to cool toward a small disk 
region.  If this is the case, the cooler gas can be dynamically 
uncoupled (or weakly coupled) with the hotter gas halo/flow.  Such a 
geometry can also allow us to deproject the density profile of the more 
spherical gas by considering only the hot gas component within 3\arcsec.  
A rigorous test of this small cooler disk model is beyond the scope of 
this paper.

\section{Negligible X-ray Contribution from Rapidly Spun-up Stars}
\label{sec:spunupstars}

\citet{SSR12} suggest that late-type main-sequence stars spun-up in
dense environments can contribute significantly to the X-ray emission in
the Bondi region of Sgr~A*, although recent {\it Chandra}   
observations have already ruled out such a possibility at Sgr~A*   
\citep{Wan+13}.

In our spectral modeling, we have modeled the X-ray emission contributed
from the stellar component (namely CV/AB) by using the $L_X$--$L_K$ relation
(Section~\ref{sec:obs}).  
On average, this should have taken into account
most of the X-ray emission from all types of stellar components,    
within the uncertainty of the
$L_X$--$L_K$ scaling relation.  However, it has been suggested that     
the rapidly spun-up stars can
dominate the X-ray emission over all other stellar components in dense   
environments such at the Galactic nucleus.
It is therefore important to know how much of the X-ray emission within the
central region (e.g., the central 1\arcsec) of NGC~3115 can be
contributed by the these rapidly spun-up stars.

First, \citet{SSR12} estimated that the effect of X-ray emission induced
by tidal spin-up is limited to a dense region where tidal spin-up is effective.
This corresponds to the central region with
high stellar density $\gtrsim 2$--$3 \times 10^7   
M_\odot$~pc$^{-3}$ within a distance $\sim 0.06$~pc from
Sgr~A*.
In NGC~3115, the
stellar density peak is about $2 \times 10^6 M_\odot$~pc$^{-3}$
\citep{EDB99} which is much smaller than the condition in Sgr~A*.
Therefore, tidal spin-up should not be important in the nuclear region
of NGC~3115.

Second, the X-ray luminosity in 2--8~keV within $\sim 0.06$~pc from 
Sgr~A* is $\sim 10^{33}$~erg~s$^{-1}$.  The total stellar mass 
(including non-spun-up stars) within that radius is estimated to be 
6--$8 \times 10^4 M_\odot$ \citep{SSR12}.  Assuming all of the X-ray 
emission comes from rapidly spun-up stars, the luminosity per unit mass 
is at most $1.7 \times 10^{28}$~erg~s$^{-1} M_\odot^{-1}$. For NGC~3115, 
the total stellar mass of the nuclear cluster of NGC~3115 with a stellar 
density peak of $2 \times 10^6 M_\odot$~pc$^{-3}$ and a characteristic 
radius of $\sim 2$~pc \citep{EDB99, Kor+96} is $\sim 7 \times 10^7 
M_\odot$. The lower density regions beyond $\sim$2~pc should not have 
significant number of these spun-up stars. Even if we assume that all 
the late-type main-sequence stars of the nuclear cluster in NGC~3115 
can be spun-up to a similar degree as to those around Sgr~A*, the total 
X-ray luminosity from these spun-up stars in the nuclear cluster can 
only be about $10^{36}$~erg~s$^{-1}$ in 2--8~keV.  Using the spectral 
model described in \citet{SSR12}, we converted the luminosity to the 
0.5--2~keV band and it is about $4 \times 10^{35}$~erg~s$^{-1}$, which 
is at least 50 times lower than the gas luminosity we determined within 
1\arcsec\ ($\sim 50$~pc).  To account for all the gas luminosity we 
measured, the X-ray emission efficiency in NGC~3115 would need to be 50 
times larger that that around Sgr~A*.  Increasing X-ray emission 
efficiency by a higher spin-up efficiency in NGC~3115 is unlikely given 
the lower stellar density at the center of NGC~3115.  Therefore, we 
conclude that spun-up stars are very unlikely to contribute 
significantly to the X-ray emission in the Bondi region of NGC~3115.

\section{Implications for Accretion Models}
\label{sec:models}

\subsection{Influence of the Black Hole}
\label{sec:bh}

\begin{figure}
\vskip 5mm
\includegraphics[width=2.8in, angle=270]{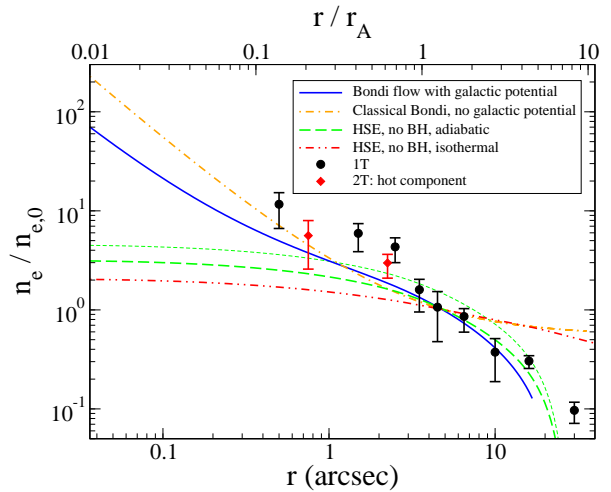}
\caption{
Density profiles of different models.  
Bondi-like flow in a galactic potential is shown in solid blue.  
The classical Bondi solution with 
the black hole potential only is shown in dot-dashed orange.  
Hydrostatic models 
without a black hole are shown in thick dashed green and dot-dot-dashed red for 
adiabatic and isothermal models, respectively.
The density 
profiles are normalized to density $n_{e,0}$ at a characteristic radius of 5\arcsec.  
The characteristic radius here is defined as $r_A \equiv 
2GM_{1.5}/c^2_{0.3}$, where $M_{1.5} \equiv 1.5 \times 10^9 M_{\odot}$ 
and $ c_{0.3}$ is 
the adiabatic sound speed at 0.3~keV.  Density profiles measured from 
the X-ray data are shown as black circles for the single temperature model and 
as red diamonds for the hot component of the two-temperature model. 
The error bars are at the 90\% 
confidence level.  Note that the error bars of the two-temperature model 
should be underestimated
because of the uncertainties in temperature.
Thin dashed green line is the adiabatic HSE model normalized at the 
measured upper density limit at 5\arcsec.
}
\label{fig:models}
\end{figure}

Since the dynamical time scale (sound crossing time) is much shorter than 
the heating, cooling, or conduction time scales near the Bondi radius 
\citep{SWI+13}, in the absence of a (supermassive) black hole, the hot 
gas should be in hydrostatic equilibrium (HSE) with the galactic 
potential.  We tested whether the measured gas density profile is consistent 
with hot gas in HSE with the galactic potential without a black hole.  
We model the hydrostatic gas density profile with the total stellar mass profile 
described in \citet{SWI+13}.  The gas mass is neglected as it is much 
smaller than the stellar mass, and dark matter is also neglected near the 
Bondi scale in which we are interested.  
Figure~\ref{fig:models} shows the density predicted by the adiabatic 
model (thick dashed green) and the isothermal model (dot-dot-dashed red).
Note that the measured temperature profile 
of the cooler component of the two-temperature model
is perhaps fairly isothermal and
the entropy of the hotter component 
is close to adiabatic\footnote{Entropy can be 
usefully quantified as $T/n_e^{2/3}$.  Since the measured temperature of 
the hotter component scales with radius as $T \sim r^{-a}$, with 
$a\lesssim 1$, and $n_e \sim r^{-1}$, the entropy 
profile is therefore close to adiabatic.}
justify the comparison 
to isothermal and adiabatic HSE models.
The HSE profiles, as well as the measured data points, are normalized 
at 5\arcsec.  
A single temperature model of the hot gas, which roughly 
corresponds to the cooler thermal component of the two-temperature 
model, is not consistent with the two HSE models, even accounting for the 
possible normalization (density) uncertainty at 5\arcsec\ (thin dashed green).  
The hot 
component of the two-temperature model is not consistent with the isothermal 
HSE model, as expected from the rising temperature toward the center.  
It is also not very consistent with the adiabatic HSE model.  
If we account for 
the normalization (density) uncertainty by shifting up the adiabatic HSE 
model by the density uncertainty at 5\arcsec\ (thin dashed green), 
the inner region is more 
consistent with the adiabatic HSE model but the density in the region 
between 5\arcsec\ and 10\arcsec\ deviates more from the model.

The classical Bondi flow model without a galactic potential is shown as a 
dot-dashed orange line in Figure~\ref{fig:models}.  
Compared to the HSE models, the 
inner most region fits better but is still inconsistent with the data.  
There is a large discrepancy beyond $\sim 8\arcsec$, due to the 
ignorance of the galactic potential in this model.  In fact, the total enclosed 
stellar mass at about 1\arcsec--2\arcsec\ is reaching $\sim 10^9 
M_{\odot}$, and therefore the galactic potential has to be taken into 
account self-consistently.  We demonstrate the effect of the galactic 
potential with a Bondi-like modeling including the galactic 
potential with the flow rate fixed to be the classical 
Bondi accretion rate (solid blue line in Figure~\ref{fig:models}).
A more realistic model 
with stellar feedback and conduction is presented in \citet{SWI+13},
although 
the existence of a billion solar mass black hole has to be an assumed prior
in that model.
Our model here is more consistent with the data in a larger radial range 
($\lesssim 10\arcsec$) compared to the HSE and the classical Bondi 
models.  In particular, the model agrees quite well with the hot 
component of the two-temperature, although the uncertainties of the gas 
density are underestimated 
because the uncertainties in temperature of the hotter component were not 
taken into account
(Section~\ref{sec:den}).  There is still 
a large discrepancy between the single temperature model (or the cooler 
component of the two-temperature model).

Our data suggest that adiabatic or isothermal HSE 
with the absence of a black hole is ruled out.  
Rotation of hot gas would give an even flatter density profile.  The 
short dynamical time scale argues against non-HSE with the absence of 
a black hole.  The X-ray data alone suggest that the rise in density 
toward the Bondi radius is more likely due to the gravitational 
influence of the supermassive black hole, in which the existence 
is supported by optical 
observations \citep{KR92,Kor+96,EDB99}.  Note that a similar technique 
has also been used to detect a massive black hole in the giant 
elliptical galaxies NGC~4649 with its Bondi radius of $\sim 1\arcsec$ 
using X-ray data alone \citep{HBB+08}.  The gravitational influence of 
the black hole should, however, compress and heat the hot gas at the 
center.  It is puzzling that the single temperature profile shows a 
decrease in temperature toward the center.  As discussed in 
Sections~\ref{sec:2T}, \ref{sec:den}, \ref{sec:SB}, and 
\ref{sec:soft}, the softer component might be located more in a small 
disk region and the hotter component of the two-temperature model might 
be the more spherical accretion/outflow component.

The X-ray data alone support that
we are witnessing the onset of an accretion (out)flow due to the 
gravitational influence of the billion solar mass supermassive black 
hole. The hot gas is likely to be in transition from the ambient gas in 
the galactic potential near the Bondi radius. 
Within $\sim 1\arcsec$--2\arcsec, the galactic potential 
becomes negligible.  Combining detailed theoretical modeling 
\citep[e.g.,][]{SWI+13} together with deeper radio observations to 
constrain the plasma properties in the vicinity of the black hole should 
allow us to distinguish among different accretion models.

Our results also suggest that in all other hot accretion flows, the 
transition regions (around the Bondi scale) connecting the ambient gas 
and the asymptotic flow near the black hole should contribute 
significant X-ray emission.  Theoretical models should take into account 
such a transition region in order to probe the inner most accretion 
region \citep[e.g.,][]{QN00}.

\subsection{Self-similar Arguments}
\label{sec:ss}

Recently, \citet{YWB12} performed simulations 
of hot accretion flows which span a much larger 
dynamical range compared to many previous simulations.  They found 
that all their numerical simulations show single power law self-similar 
profiles spanning close to the Bondi radius down to 1--$10 R_S$.  Such 
results agree with many other previous simulations.  In particular, they 
found that all radial profiles scale with similar power law indexes 
regardless of viscosity, magnetic field, or initial conditions.  
For instance, their simulated 
density profile scales as $\rho \propto r^{-3/2+p}$, with $p=0.65$--0.85 
from their simulations\footnote{Note the different symbols defined for 
the scaling relations in \citet{YWB12}.}, which is consistent with our 
measured density profile of $\rho \propto r^{-[0.62^{+0.26}_{-0.38}]}$ 
within 3\arcsec\ (141~pc) for our single temperature model and close to 
$\rho \propto r^{-1}$ for a wider range of radii and for the hotter 
component of the two-temperature model. 
Hot gas near the Bondi radius may still be in transition from the 
ambient ISM to the accretion flow, and therefore the density slope near 
the Bondi radius may not reach the asymptotic value of the accretion 
flow \citep[e.g.,][]{Bon52,Qua02}.  For accretion flow in a galactic 
potential, the transition may be smoother than the accretion flow in a 
uniform ambient ISM \citep[e.g.,][]{QN00}.
If the asymptotic density profile 
toward the black hole is close to the predictions by \citet{YWB12}, 
the density profile of NGC~3115 may be more smoothly in transition from the 
region around the Bondi radius all the way toward the event horizon, and 
therefore a single power-law in density between $\sim$$R_S$ and $\sim$$R_B$ 
may be applicable.

For comparison, \citet{Wan+13} recently estimated a density power law 
index of one for Sgr~A*, which generally agrees with NGC~3115.  Since the 
spectrum in Sgr~A* is not spatially resolved, they had to assume a 
temperature profile of the form $T \propto r^{-1}$ to estimate the density 
profile.  Similar to NGC~3115, the flow in Sgr~A* may also be in 
transition near its Bondi radius so that the asymptotic $T \propto 
r^{-1}$ behavior may not be applicable.  
A flatter temperature profile would give 
a flatter density profile which is more consistent with the density 
profile of our single temperature model in NGC~3115.

At about 1\arcsec\ (47~pc), the electron density of the single 
temperature model is 0.15~cm$^{-3}$.  While there are larger 
uncertainties in the two-temperature model and the metallicity, the 
density should not be off by more than an order of magnitude.  The 
accretion rate at 1\arcsec\ (47~pc) is then estimated to be $\dot M_{\rm 
acc}(47~{\rm pc}) = 4 \pi \lambda R^2 \rho c_s = 9 \times 10^{-3} 
M_{\odot} {\rm ~yr}^{-1}$, where $\lambda = 0.25$ and $\gamma = 5/3$ for 
an adiabatic process, $c_s = \sqrt{\gamma k_B T/\mu m_p}$ is the 
adiabatic sound speed, $\mu = 0.63$ is the mean molecular weight, and $T 
= 0.3$~keV is assumed.  The uncertainty of $T$ also will not introduce 
an error in the accretion rate by more than an order of magnitude.

The mass accretion rate estimated at 1\arcsec\ (47~pc) from the black 
hole is a factor of a few smaller than the accretion rate of (2--$4) 
\times 10^{-2} M_{\odot}$~yr$^{-1}$ estimated around a larger radius of 
4\arcsec--5\arcsec.  This highlights the systematic uncertainty in estimating 
the accretion rate near or beyond the Bondi radius in other galaxies 
with unresolved/underresolved Bondi radii.
The estimation near 
1\arcsec\ (47~pc) is closer to the upper limit of $2 \times 10^{-3} 
M_{\odot}$~yr$^{-1}$ we estimated in a more self-consistent model 
\citep{SWI+13}.

The upper limit of the X-ray luminosity of the central AGN is $4.4 
\times 10^{37}$~erg~s$^{-1}$, which is about six orders of magnitude 
smaller than the accretion luminosity ($5\times 10^{43}$~erg~s$^{-1}$) 
at 1\arcsec\ if we assume a 10\% radiative efficiency.  As discussed in 
W11, this discrepancy can be explained if the accretion rate near 
the black hole is suppressed as predicted by ADIOS or CDAF models with 
the scaling relation $\dot M \propto r^p$.  \citet{YWB12} found in their 
simulations that the accretion rate is constant within about $10 R_S$ 
and the accretion rate scales as $\dot M \propto r^p$ beyond $10 R_S$, 
as expected in the ADIOS model.  Assuming the accretion rate at $10 R_S$ is 
suppressed by six orders of magnitude, the scaling relation gives 
$p\approx 1.3$ for a billion solar mass black hole.  Such a high value 
of $p$ is outside the theoretical upper limit of 1 and also larger than 
most of the simulated results of 0.5--0.7 \citep{YWB12}.  However, 
observational uncertainties may bring $p$ close to one, which is consistent 
with the latest version of the ADIOS model \citep{Beg12}.  Such a high 
$p=1$ value suggests a very flat density profile of $\rho \propto 
r^{-0.5}$, consistent with the power law slope of $0.62^{+0.26}_{-0.38}$ 
(90\% confidence) within 3\arcsec\ (141~pc) for the single temperature 
model.

It is also likely that other factors are needed to explain the large 
discrepancy in X-ray luminosity and accretion rate.  For example, 
accretion may be highly suppressed by rotation close to the event 
horizon of the black hole \citep{PB03,LOS13} and significant outflow can 
also be generated during the rotational accretion process 
\citep{BB99,LOS13}, although the suppression can be less effective if 
the gas viscosity is sufficiently high \citep{NF11}.  Numerical 
simulations show that rotation can flatten density and temperature 
profiles \citep{BMH+09}. 
Stellar feedback should also suppress the accretion as
discussed in \citet{HS13} \& \citet{SWI+13}, with the latter authors also 
including conduction as a possible suppression mechanism.
It is also possible that the radiation 
efficiency can be lower than the 10\% canonical value we assumed 
\citep{Ho08}.

\subsection{Feedback Models}
\label{sec:feedback}

The discussion above was based on steady or quasi-steady state flows 
without feedback. 
However, the dynamics of hot gas within a Bondi radius is not only 
governed by the black hole (and the galactic) potential alone.  If there 
is (was) a sudden release of feedback energy from the black hole 
(recently) as we see in other giant elliptical galaxies, dynamical 
disturbance will be important \citep[e.g.,][]{Fab12,MN12}.  Since there 
is no strong evidence of AGN feedback in NGC~3115 (e.g., strong radio 
source, jet, or X-ray bubble), we do not consider strong dynamical flow 
in this paper.  For weak AGN such as NGC~3115 or Sgr~A*, other feedback 
mechanisms such as stellar feedback or conduction can also play  
important roles \citep{HS13,SAG+13,SWI+13}.  Therefore, resolving the gas 
profiles within the Bondi radius is only a step further toward the 
understanding of black hole accretion.  Realistic theoretical modeling 
and simulations should be performed to match observations, and we 
present our effort by including conduction and stellar feedback in 
NGC~3115 in our companion paper \citep{SWI+13}.

\section{Summary and Conclusions}
\label{sec:summary}

With a
temperature of 0.3~keV for the ambient hot gas, the Bondi radius
of the supermassive black hole in NGC~3115 is $R_B 
= 112$--224~pc = 2\farcs4--$4\farcs8$ (W11).  
Radio observations have recently detected a weak AGN with $L_{\rm 8.5 GHz} 
= 3.1 \times 10^{35}$~erg~s$^{-1}$ at the galaxy center \citep{WN12}.  
We searched for a signature of the AGN in X-ray but we did not find any 
strong evidence of a central point source.  We determined the upper 
limit of the X-ray luminosity to be $L_{X, \rm AGN} = 4.4\, (1.1) 
\times 10^{37}$~erg~s$^{-1}$ in 0.5--6.0 (0.5--1.0)~keV.  The Eddington 
fraction is thus $L_{X, \rm AGN}/L_{\rm Edd} < 3.5 \times 10^{-10} (10^9 
M_{\sun}/M_{\rm BH})$, making it one the most under-luminous AGNs 
\citep{Ho08}.  Therefore, the accretion of the NGC~3115 black hole is 
expected to be in the hot mode with an expected temperature profile 
close to the virial temperature of the system and increasing toward the 
center as $T \propto r^{-1}$ \citep[see, e.g.,][and references therein]{NM08}.

The hot gas component of the X-ray emission within the Bondi radius is 
clearly extended and is resolved both spatially and spectrally.  The hot 
thermal plasma is robustly detected out to $\sim 10 R_B$ (a few tens of 
arcsec).  We studied accretion-model independent temperature and density 
profiles within and around the Bondi radius.

The projected temperature of a single temperature model of the ambient 
hot gas is slowly increasing from the outer region of $\sim 
30\arcsec$--40\arcsec\ ($\sim$1.5--2~kpc) toward 5\arcsec\ (235~pc), 
consistent with 0.3~keV.  The projected temperature jumps significantly 
to a higher temperature of $\sim 0.7$~keV within $\sim 
4\arcsec$--5\arcsec, but then abruptly drops back to $\sim 0.3$~keV 
within $\sim 3$\arcsec\ (141~pc).  This conflicts with the 
theoretical expection that the temperature should be rising toward the 
center, suggesting that there is significant softer emission within a 
scale of $\sim 150$~pc (around the Bondi scale) compared to a simple hot 
accretion model with a monotonic increase in temperature.

With the high quality Megasecond {\it Chandra} data, 
we found evidence that at least a 
two-temperature model is needed in the inner few arcsec (150~pc).  The 
hotter temperature of the two-temperature model increases toward the 
center to $\sim 1$~keV, consistent with predictions from hot accretion models.
The softer component, which dominates over the hotter component in 
emission measure (gas normalization) by a factor of two to four, has a 
temperature of $\sim 0.3$~keV.
The softer component 
cannot be accounted for by projection of cooler surrounding gas with a spherically 
symmetric distribution.  Even if we assume the cooler surrounding gas 
is distributed as a pancake-like ellipsoid roughly following the optical 
light or a very thick disk structure, this can at most account for about 
22\% of the softer component in the central 1\arcsec--3\arcsec.  
We argued that 
the cooler component at the center is indeed physically located in the 
central $\sim$150~pc rather than projected gas from the outer 
region, unless the distribution of the outer cooler gas preferentially 
aligns toward the line-of-sight of the supermassive black hole, which is 
not very likely.
Tidally-stripped cores of giant stars near the supermassive back hole may 
emit soft X-rays \citep{DGM+01,DK05}, 
but it is also unlikely to explain most of the softer X-ray emission.

We argued that the softer component in the central 150~pc more likely 
comes from diffuse gas which can be in a multi-temperature phase, as 
supported by recent numerical simulations \citep[e.g.,][]{GRO13}.  
We also noticed 
some weak evidence that the cooler component is preferentially located 
along the major axis, resembling a small thin disk seen on an optical 
{\it HST} image \citep{Kor+96}. The hotter component is more spherically 
distributed.  The cooler component may be circulating and cooling toward a 
disk region.  

\citet{SSR12} suggested that late-type main-sequence stars spun-up in 
dense environment can contribute significantly to X-ray emission in 
galactic centers.  We argued that it is very 
unlikely to be the case in NGC~3115.

The density profile of NGC~3115 suggests that hot gas in 
adiabatic or isothermal HSE with the 
galactic potential in the absence of a black hole is ruled out.  The 
short dynamical time scale also argues against non-HSE without a black 
hole.  Therefore, we are witnessing the onset of an accretion 
(out)flow influenced by the strong gravity of the supermassive black 
hole.  It is puzzling, however, that the single temperature profile 
drops at the center rather than compressionally heated to a higher 
temperature, although the detected hot gas may be in a multi-temperature 
phase as mentioned above.

We determined that the density profile is broadly consistent with $\rho 
\propto r^{-1}$ within 5\arcsec\ (235~pc) around the Bondi radius for 
either the single temperature or the two-temperature model.  In 
particular, the density profile flattens to $\rho \propto 
r^{-[0.62^{+0.26}_{-0.38}]}$ within 3\arcsec\ (141~pc) for the single 
temperature model.  This is remarkably consistent with the narrow range 
of power law index of 0.65--0.85 determined from a large number of 
numerical simulations spanning a very large dynamical radial range 
\citep{YWB12}.
Note that the density we determined depends on the assumed geometry.  If 
the gas is significantly clumpy and/or if the filling factors of 
two-temperature model are less than one, the density determined is 
overestimated.

We estimated that the accretion rate at 1\arcsec\ (47~pc) to be $\dot 
M_{\rm acc}(47~{\rm pc}) = 9 \times 10^{-3} M_{\odot} {\rm ~yr}^{-1}$, 
which is a factor of a few smaller than the accretion rate determined at 
a larger radius of 4\arcsec--5\arcsec. This illustrates the systematic 
uncertainty in estimating the accretion rate near or beyond the Bondi 
radius in galaxies for which the Bondi region is not spatially resolved.

Since the upper limit of the X-ray luminosity of the central AGN is 
about six orders of magnitude smaller than the accretion luminosity, hot 
gas actually accreted through the event horizon must be highly 
suppressed by, e.g., outflow, rotational support, and/or stellar 
feedback.  Radiation efficiency may also be much lower than the 10\% 
canonical value.

Future mission like {\it 
SMART-X}\footnote{http://smart-x.cfa.harvard.edu/} with an order of 
magnitude increase in effective area compared to {\it Chandra} will 
allow us to collect enough photons to rigorously study the dynamical 
properties of the accretion flow within the Bondi radius of NGC~3115 (and 
potentially M31*).  The sub-arcsec resolution comparable to {\it Chandra} 
is essential.  
On the other hand, combining high angular resolution 
radio observations by, e.g., the Event Horizon 
Telescope\footnote{http://www.eventhorizontelescope.org}, to probe the 
hot gas properties around the very large event horizon (2--4$\mu$as) of 
NGC~3115 should allow us to understand how gas is being accreted from 
the Bondi radius down to the black hole.

\acknowledgments

We thank the referee for the comments.  We also thank Rosanne Di 
Stefano, Eric Emsellem, Dacheng Lin, Peter Maksym, William Mathews, Paul 
Nulsen, Yuanyuan Su, and Feng Yuan for useful discussions.
The work is supported by {\it Chandra} XVP grant
GO2-13104X.
RVS is supported by NASA Hubble Fellowship grant HST-HF-51298.01.
Some of the data presented in this paper were obtained from the 
Mikulski Archive for Space Telescopes (MAST). STScI is operated by the 
Association of Universities for Research in Astronomy, Inc., under NASA 
contract NAS5-26555. Support for MAST for non-HST data is provided by 
the NASA Office of Space Science via grant NNX09AF08G and by other 
grants and contracts.

\appendix

\section{A. Systematic Uncertainties}
\label{sec:app1}

\begin{figure*}
\includegraphics[width=0.82\textwidth, angle=270]{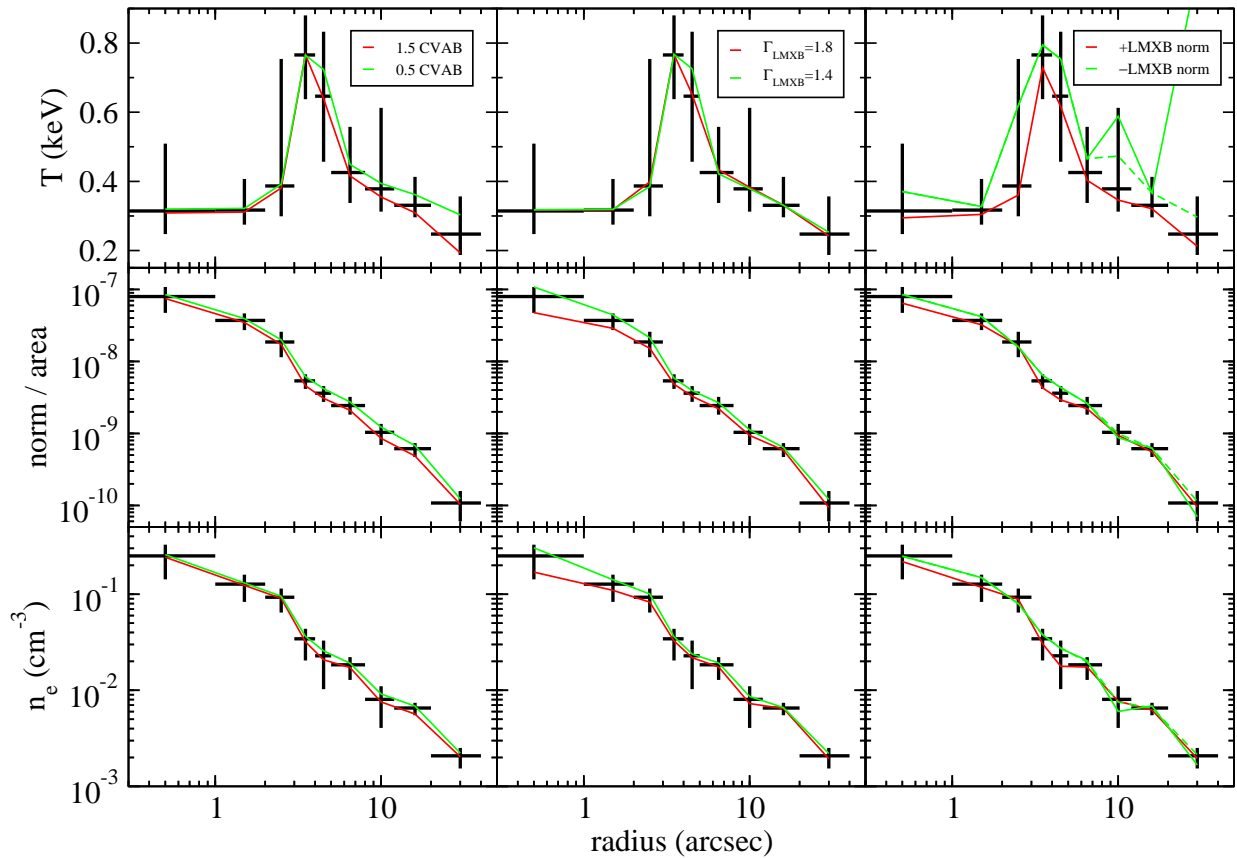}
\caption{
Left column: Systematic uncertainties introduced by CV/AB 
normalizations.  The upper panel shows the temperature profiles of the 
nominal single temperature model (black). Model with CV/AB contribution 
increased (decreased) by 50\% is shown in red (green).  The middle panel 
shows the {\tt APEC} normalization per unit area 
(cm$^{-5}$~pixel$^{-2}$) of the nominal single temperature model (black) 
and models with CV/AB contribution changed by +50\% (red) and -50\% 
(green).  The lower panel shows the deprojected density profiles of the 
nominal single temperature model (black) and models with CV/AB 
contribution changed by +50\% (red) and -50\% (green).\\ 
Middle column: 
Similar as left row, but with the red (green) lines representing a model with 
LMXB power-law index of $\Gamma_{\rm LMXB} = 1.8 (1.4)$.\\ 
Right column: 
Similar as left row, but with the red (green) lines representing a model with 
LMXB normalizations fixed to their upper (lower) limits (90\% 
confidence).  Note that there are local minima in c-statistics within 
the 90\% confidence regions for the model with LMXB normalizations fixed 
to the lower limits, and the profiles corresponding to these local minima 
are shown as dashed green lines.\\ For all panels, vertical error bars 
are at the 90\% confidence level and horizontal bars indicate the radial 
binning size.
}
\label{fig:sys_set1}
\end{figure*}

\begin{figure*}
\includegraphics[width=0.82\textwidth, angle=270]{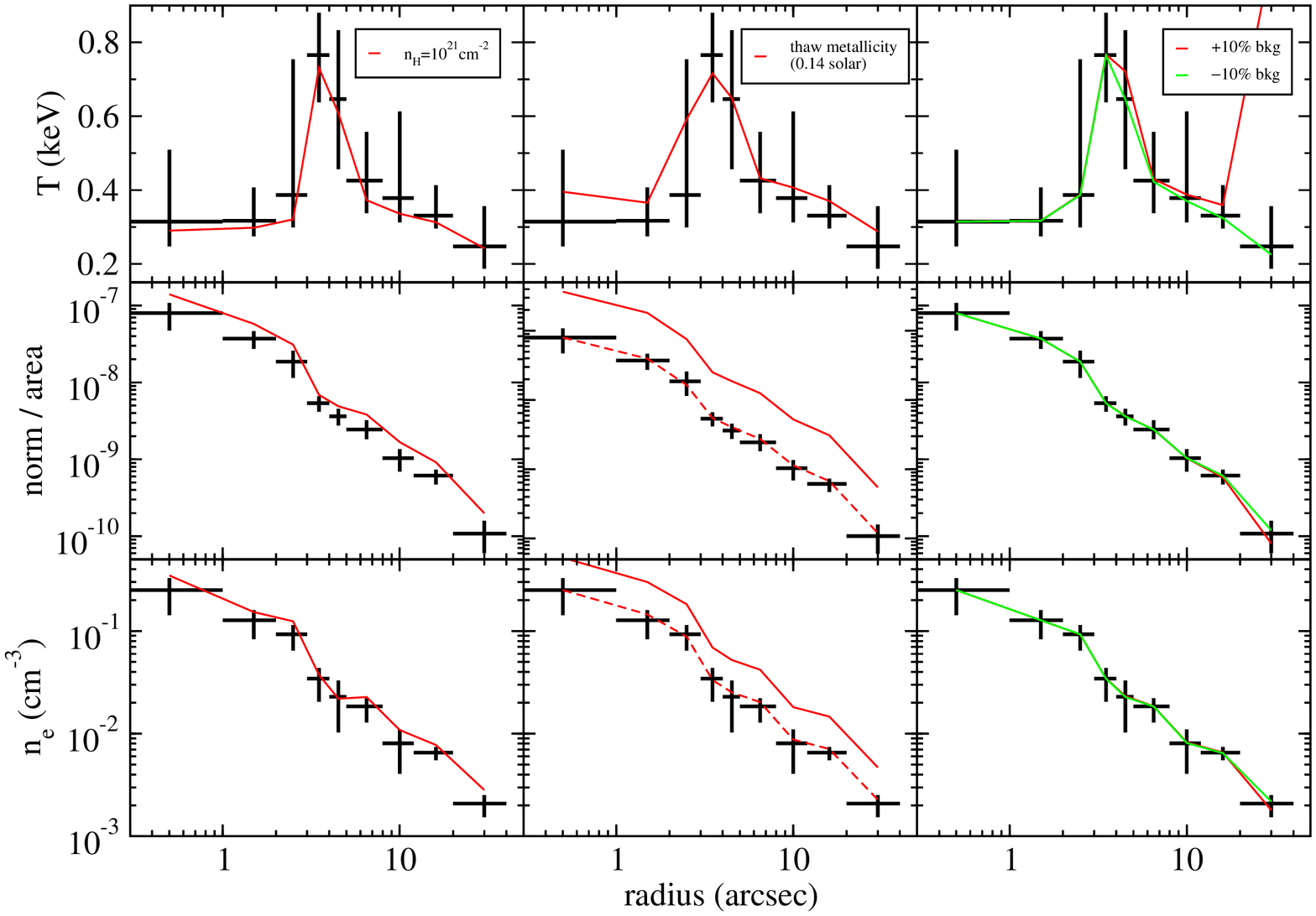}
\caption{
Left column: Similar as the left row in Figure~\ref{fig:sys_set1}, but 
with the red lines representing a model with higher absorption ($n_{\rm H} = 
10^{21}$~cm$^{-2}$).\\ 
Middle column: Similar as the left row, but with 
the red lines representing a model with abundance thawed.  The best-fit 
metallicity is 0.14 solar. 
The red dashed lines in the middle and lower panels represent the same 
model with abundance thawed but normalized to the nominal model (black) 
at the central bin.
Note that in the middle row panel, 
the y-axis has a scale slightly different from the other columns.  The 
major tick marks corresponds to those of the other columns.\\ 
Right 
column: Similar as the left row, but with the red (green) lines 
representing a model with local background level changed by +10 (-10)\%.
}
\label{fig:sys_set2}
\end{figure*}

In Section~\ref{sec:1T}, we characterized the projected spectrum of the 
hot gas component by a single temperature model.  Although the single 
temperature model may not be a physically correct model to describe the 
thermal plasma, it generally gives good fits to characterize the 
projected spectra.  To ensure this single temperature characterization 
of the projected spectra is robust, we checked against systematic 
uncertainties in spectral modeling extensively as outlined below.

\subsection{CV/AB uncertainties}
\label{sec:sys_cvab}

We modeled the CV/AB contribution in X-ray by assuming the $L_X$--$L_K$ 
scaling relation determined from M32.  The galaxy by galaxy variation is about 30\% 
(rms) for 11 nearby early-type galaxies \citep{BG11} and about 30--40\% 
for 3 nearby early-type galaxies studied by \citet{RCS+08}.  We varied the CV/AB 
normalizations by $\pm 50\%$ which is comparable to these galaxy by 
galaxy variations and also larger than the statistical uncertainties of 
about 30\% (90\% confidence level) in the spectral fitting of M32.  
Varying the CV/AB normalizations only introduces very small systematic 
biases to the temperatures or gas normalizations (densities) within 
$\sim 10\arcsec$ and larger ones beyond that (left column in 
Figure~\ref{fig:sys_set1}).  All of the systematic uncertainties are 
within the statistical uncertainties (90\% confidence).

\subsection{Unresolved LMXB uncertainties}
\label{sec:sys_lmxb}

The power-law index of the combined spectrum of all the resolved point 
sources within $D_{25}$ is $\Gamma_{\rm LMXB} = 1.61^{+0.02}_{-0.02}$, 
consistent with the value of 1.6 determined in the bulge of M31 
\citep{IAB03}.  We assessed the systematic uncertainties of the 
power-law index of unresolved LMXBs by changing $\Gamma_{\rm LMXB}$ to 
1.4 and 1.8 (middle column in Figure~\ref{fig:sys_set1}). Changing the 
LMXB power-law index to 1.8 only introduces negligible systematic 
uncertainties of less than 3\% in temperature.  A power-law index of 1.4 
changes the temperature by less than 2\% in general, but with a large 
change of 12\% in the 4\arcsec--5\arcsec\ bin.  These are all well 
within the 90\% confidence of the statistical uncertainties.  The 
normalizations are generally changed by 10--20\% and at most by about 
40\%.  This generally leads to density biases of less than 10\% and at 
most by $\sim 20$--$30\%$. All of these are comparable to the 
statistical uncertainties.

We also examined the systematic uncertainties of the normalizations of 
the unresolved LMXBs by fixing them to their upper and lower limits (90\% 
confidence) of the nominal fitting (right column in 
Figure~\ref{fig:sys_set1}). Fixing the LMXB normalizations to their 
lower limits (solid red) generally lowers the gas temperatures and 
normalizations (densities), but within the statistical uncertainties. 
When we fixed the LMXB normalizations to their lower limits (solid 
green), the temperatures are biased higher, particularly in the outer 
regions.  We noticed that in the seventh and the ninth bins, there are 
local minima with deviations of c-statistics from the global minimum 
$\Delta C < 1$.  The profiles at these local minimums (dashed green 
lines) are much closer to the nominal profiles.  These larger biases in 
temperatures in the outer regions may be due to the underestimation of 
the hard photons from LMXBs and therefore the thermal component in the 
model tries to compensate for the excess hard photons with higher 
temperatures.  The existence of the second minimum indicates the 
significance of a thermal component that is closer to the nominal model.  
Nevertheless, the gas normalization and density profiles are hardly 
biased by the systematic uncertainties of the unresolved LMXB 
normalizations.

\subsection{Hydrogen column density uncertainties}
\label{sec:sys_nH}

Ideally, the absorption should be fitted from the spectrum, but thawing the 
absorption gives an unphysically low $n_{\rm H}$ of zero, suggesting a 
degeneracy between $n_{\rm H}$ and the hot gas component.  Given the 
generally low $n_{\rm H}$ intrinsic to early-type galaxies and also the 
insignificant amount of total HI (less than a few $10^7 M_{\odot}$) in 
NGC~3115 \citep{RHB+91,KKH+04,SW06}, we fixed $n_{\rm H} = 4.32 \times 
10^{20}$~cm$^{-2}$ to the Galactic value \citep{DL90}.  Fitting the 
$n_{\rm H}$ with bright resolved LMXB in NGC~3115 generally gives a 
higher $n_{\rm H}$ by at most $60\%$ but consistent with the Galactic 
value within $1\sigma$--$3\sigma$ in uncertainties.  We assessed the 
systematic uncertainty in absorption by fixing $n_{\rm H} = 10 \times 
10^{20}$~cm$^{-2}$ (left column in Figure~\ref{fig:sys_set2}).  Even if 
the absorption is this high, the temperatures are only slightly biased 
lower and the gas normalizations (densities) are biased higher to the 
upper limit of the uncertainties of the nominal model.  This does not 
change our results qualitatively.

\subsection{Metallicity uncertainties}
\label{sec:sys_abun}

Fitting the metallicity of the hot gas from X-ray spectra gives a very 
low abundance of 0.14 solar.  However, this sub-solar abundance is not 
expected since the hot gas should be contributed by stellar feedback 
which should give solar or super-solar abundance \citep[see,][for a more 
detailed discussion]{SWI+13}.  It is also known that metallicity 
determination can be biased low, particularly due to low spectral 
resolution of ccds and multi-temperature structure of the hot gas 
\citep{Buo00,SI13}. Motivated by a more realistic physical situation, we 
fixed the abundance to the solar value.  At temperatures below 1~keV, the 
emission is dominated by line emission which is proportional to 
metallicity.  This introduces a degeneracy between metallicity and gas 
density (because emission is also proportional to density squared).  
Thawing the abundance only introduces a systematic uncertainty in 
temperature that is smaller than the statistical uncertainty and 
increases the gas normalizations by a factor of four to five (middle 
column in Figure~\ref{fig:sys_set2}). The derived density only increases 
by a factor of two without affecting the density slope.  All our 
conclusions remain the same.

\subsection{Background uncertainties}
\label{sec:sys_bkg}

The background only contributes from less than about 1(2)\% of the 
0.5--2.0 (0.5--6.0~keV) emission at the central 3\arcsec\ up to 
6(10)\% at 8\arcsec.
The background increases to about 50\% (60--70\%) 
of the emission in the 0.5--2.0 (0.5--6.0)~keV energy band 
between 20\arcsec--40\arcsec. 
We changed the background level by $\pm 
10\%$.  This generally introduces less than 2\% systematic uncertainties 
in temperature within $\sim 10\arcsec$, but larger beyond that (right 
column in Figure~\ref{fig:sys_set2}).  The temperature beyond 20\arcsec\ 
can be significantly biased.
Nevertheless, 
the gas normalizations is biased by less than 1\% within 
$\sim 10\arcsec$ and at most 25\% in the outermost bin (but still 
smaller then its statistical uncertainty).  The bias in the density profile 
is even smaller.  Note that the outermost regions only contribute a 
small fraction of emission to the projected emission of the inner 
regions.  Therefore, the deprojected density profile of the inner 
regions is not sensitive to the precise value of the density in the 
outermost regions.

\section{B. X-ray Limits of the Weak AGN}
\label{sec:appAGN}

We assessed a conservative upper limit of the potential central 
point source by modeling the spatial distribution of the X-ray emission in 
5\arcsec\ with a two-component model: a point source component and an 
extended diffuse component.  We model the point source surface 
brightness profile with a Moffat model.
The Moffat model 
is of the form $S = S_0 [1+(r/r_c)^2]^{-n}$, where $r_c$ and $n$ are 
fixed to the best-fit values from a nearby point source, and $S_0$ is a free 
parameter which characterizes the contribution of the potential point 
source.  We model the extended diffuse component with another Moffat 
model and thawed all its three parameters.  The best-fit model of the 
two-component model (Moffat+Moffat) is shown in the right panel of 
Figure~\ref{fig:psfcompare}.  For comparison, we also fitted a single 
Moffat model (no point source model) and it is plotted on the same 
figure.  We find that a single Moffat model gives $\chi^2 = 146$ with 97 
degrees of freedom for the 0.5--6.0~keV band.  The two-component model 
(Moffat+Moffat) gives $\chi^2 = 128$ with 96 degrees of freedom.  A 
simple F-test gives an F-statistics of 13.5 with a probability of $4 
\times 10^{-4}$, strongly suggesting that the two-component model is 
preferred (or a point source is present).  This two-component model 
suggests that 19-31\% (90\% confidence interval) of the diffuse emission within 
a radius of 1\arcsec\ region in 0.5--6.0~keV comes from the point source.  
Note that the Moffat model has a flat core, and therefore modeling 
the extended emission with this model may underestimate the extended 
emission at the center if the true extended profile is more sharply 
peaked.  Therefore, our estimation of the point source 
contribution should be regarded as an upper limit.  

By performing 
similar analysis in the soft (0.5--1.0~keV), medium (1.0--2.0~keV), and 
hard (2.0--6.0~keV) energy bands, we have constrained 
the photon counts in each of these energy bands from the potential point 
source.  We fit a {\tt PHABS*POWERLAW} model using {\tt XSPEC} to these 
three energy bands with the absorption fixed at the Galactic value as before.  
We determined that the best-fit two-component model gives a power-law 
index of 2.2 with an aperture corrected absorbed luminosity of $2.9\, 
(0.9) \times 10^{37}$~erg~s$^{-1}$ in 0.5--6.0 (0.5--1.0)~keV.  Using 
the upper limits of the two-component model gives a power-law index 
of 2.0 with an aperture corrected absorbed luminosity of $4.4\, (1.1) 
\times 10^{37}$~erg~s$^{-1}$ in 0.5--6.0 (0.5--1.0)~keV.

When we included these spectral models for the potential point source in the 
spectral fitting within a circular region of 1\arcsec\ in radius, 
this only increases the best-fit temperature of hot gas 
(Section~\ref{sec:1T}) by at most 23\% compare to a model without an 
AGN.  Such an increase is much smaller than the statistical uncertainty 
(90\% confidence interval or 95\% one-sided uncertainty).  The best-fit 
gas normalization (flux) is at most lowered by about 40\%, which is 
at the lower limit of the 90\% confidence interval.  Since our model is 
conservative, we conclude that the potential AGN should not contribute 
much to the X-ray emission.


\begin{thebibliography}{}

\bibitem[{{Abramowicz et al.}(2002) {Abramowicz}, {Igumenshchev}, {Quataert}, \& {Narayan}}]{AIQ+02}
{Abramowicz}, M.~A., {Igumenshchev}, I.~V., {Quataert}, E., \& {Narayan}, R. 2002, \apj, 565, 1101

\bibitem[{{Baganoff et al.}(2003)}]{Bag+03}
{Baganoff}, F.~K.., et al. 2003, \apj, 591, 891


\bibitem[{{Barai et al.}(2012){Barai}, {Proga}, \& {Nagamine}}]{BPN12}
{Barai}, P., {Proga}, D., \& {Nagamine}, K. 2012, \mnras, 424, 728

\bibitem[{{Begelman}(2012)}]{Beg12}
{Begelman}, M.~C. 2012, \mnras, 420, 2912

\bibitem[{{Blandford \& Begelman}(1999)}]{BB99}
{Blandford}, R.~D., \& {Begelman}, M.~C. 1999, \mnras, 303, L1

\bibitem[{{Bogd{\'a}n \& Gilfanov}(2011)}]{BG11}
{Bogd{\'a}n}, {\'A}., \& {Gilfanov}, M. 2011, \mnras, 418, 1901

\bibitem[{{Bondi}(1952)}]{Bon52}
Bondi, H. 1952, \mnras, 112, 195

\bibitem[{{Boroson et al.}(2011) {Boroson}, {Kim}, \& {Fabbiano}}]{BKF11}
{Boroson}, B., {Kim}, D.-W., \& {Fabbiano}, G. 2011, \apj, 729, 12

\bibitem[{{Brighenti \& Mathews}(1999)}]{BM99}
{Brighenti}, F., \& {Mathews}, W.~G. 1999, \apj, 527, L89

\bibitem[{{Brighenti et al.}(2009){Brighenti}, {Mathews}, {Humphrey}, \& {Buote}}]{BMH+09}
{Brighenti}, F., {Mathews}, W.~G., {Humphrey}, P.~J., \& {Buote}, D.~A. 2009, \apj, 705, 1672

\bibitem[{{Buote}(2000)}]{Buo00}
{Buote}, D.~A. 2000, \mnras, 311, 176

\bibitem[{{Buote et al.}(2003){Buote}, {Lewis}, {Brighenti}, \& {Mathews}}]{BLB+03}
{Buote}, D.~A., {Lewis}, A.~D., {Brighenti}, F., \& {Mathews}, W.~G. 2003, \apj, 594, 741

\bibitem[{{Das \& Sharma}(2013)}]{DS13}
{Das}, U., \& {Sharma}, P. 2013, \mnras, 435, 2431

\bibitem[{{Davies} \& {King}(2005)}]{DK05}
{Davies}, M.~B., \& {King}, A. 2005, \apj, 624, L25

\bibitem[{{Dickey} \& {Lockman}(1990)}]{DL90}
{Dickey}, J.~M., \& {Lockman}, F.~J. 1990, \araa, 28, 215

\bibitem[{{Di Stefano et al.}(2001){Di Stefano}, {Greiner}, {Murray}, \& {Garcia}}]{DGM+01}
{Di Stefano}, R., {Greiner}, J., {Murray}, S., \& {Garcia}, M. 2001, \apjl, 551, L37

\bibitem[{{Dressler \& Gunn}(1983)}]{DG83}
{Dressler}, A., \& {Gunn}, J.~E. 1983, \apj, 270, 7

\bibitem[{{Emsellem et al.}(1999){Emsellem}, {Dejonghe}, \& {Bacon}}]{EDB99}
{Emsellem}, E., {Dejonghe}, H., \& {Bacon}, R. 1999, \mnras, 303, 495

\bibitem[{{Fabian}(2012)}]{Fab12}
{Fabian}, A.~C. 2012, \araa, 50, 455

\bibitem[{{Fabian \& Rees}(1995)}]{FR95}
{Fabian}, A.~C., \& {Rees}, M.~J. 1995, \mnras, 277, L55

\bibitem[{{Gayley}(2013)}]{Gay13}
{Gayley}, K.~G. 2013, in American Astronomical Society HEAD Meeting, Two-temperature and Model-Independent Differential Emission Measure Distributions: The Emperor's New Clothes?, ed. American Astronomical Society, 13, 117.02

\bibitem[{{Garcia et al.}(2010)}]{Gar+10}
{Garcia}, M.~R., {Hextall}, R., {Baganoff}, F.~K., 
{Galache}, J., {Melia}, F., {Murray}, S.~S., {Primini}, F.~A., {Sjouwerman}, L.~O., \& {Williams}, B. 2010, \apj, 710, 755

\bibitem[{{Gaspari et al.}(2013){Gaspari}, {Ruszkowski}, \& {Oh}}]{GRO13}
{Gaspari}, M., {Ruszkowski}, M., \& {Oh}, S.~P. 2013, \mnras, 432, 3401

\bibitem[{{Guo \& Mathews}(2013)}]{GM13}
{Guo}, F., \& {Mathews}, W.~G. 2013, arXiv:1305.2958

\bibitem[{{Hillel \& Soker}(2013)}]{HS13}
{Hillel}, S., \& {Soker}, N. 2013, \mnras, 430, 1970

\bibitem[{{Ho}(2008)}]{Ho08}
Ho, L.~C. 2008, \araa, 46, 475

\bibitem[{{Ho}(2009)}]{Ho09}
Ho, L.~C. 2009, \apj, 699, 626

\bibitem[{{Huchra \& Burg}(1992)}]{HB92}
{Huchra}, J., \& {Burg}, R. 1992, \apj, 393, 90

\bibitem[{{Humphrey et al.}(2008){Humphrey}, {Buote}, {Brighenti}, {Gebhardt}, \& {Mathews}}]{HBB+08}
{Humphrey}, P.~J., {Buote}, D.~A., {Brighenti}, F., {Gebhardt}, K., \& {Mathews}, W.~G. 2008, \apj, 683, 161

\bibitem[{{Ichimaru}(1977)}]{Ich77}
Ichimaru, S. 1977, \apj, 214, 840

\bibitem[{{Irwin et al.}(2003){Irwin}, {Athey}, \& {Bregman}}]{IAB03}
{Irwin}, J.~A., {Athey}, A.~E., \& {Bregman}, J.~N. 2003, \apj, 587, 356

\bibitem[{{Kaastra} {et~al.}(2008){Kaastra}, {Paerels}, {Durret}, {Schindler}, \& {Richter}}]{KPD+08}
{Kaastra}, J.~S., {Paerels}, F.~B.~S., {Durret}, F., {Schindler}, S., \& {Richter}, P. 2008, SSRv, 134, 155

\bibitem[{{Karachentsev} {et~al.}(2004){Karachentsev}, {Karachentseva}, {Huchtmeier}, \& {Makarov}}]{KKH+04}
{Karachentsev}, I.~D., {Karachentseva}, V.~E., {Huchtmeier}, W.~K., \& {Makarov}, D.~I. 2004, \aj, 127, 2031

\bibitem[{{Kormendy \& Richstone}(1992)}]{KR92}
{Kormendy}, J., \& {Richstone}, D. 1992, \apj, 393, 559

\bibitem[{{Kormendy et al.}(1996)}]{Kor+96}
{Kormendy}, J., et al. 1996, \apjl, 473, L91

\bibitem[{{Kriss} {et~al.}(1983){Kriss}, {Cioffi}, \& {Canizares}}]{KCC83}
{Kriss}, G.~A., {Cioffi}, D.~F., \& {Canizares}, C.~R. 1983, \apj, 272, 439

\bibitem[{{Ledo} {et~al.}(2010){Ledo}, {Sarzi}, {Dotti}, {Khochfar}, \& {Morelli}}]{LSD+10}
{Ledo}, H.~R., {Sarzi}, M., {Dotti}, M., {Khochfar}, S., \& {Morelli}, L. 2010, \mnras, 407, 969

\bibitem[{{Li} {et~al.}(2013){Li}, {Ostriker}, \& {Sunyaev}}]{LOS13}
{Li}, J., {Ostriker}, J., \& {Sunyaev}, R. 2013, \apj, 767, 105

\bibitem[{{MacLeod} {et~al.}(2012){MacLeod}, {Guillochon}, \& {Ramirez-Ruiz}}]{MGR12}
{MacLeod}, M., {Guillochon}, J., \& {Ramirez-Ruiz}, E. 2012, \apj, 757, 134

\bibitem[{{Magorrian \& Tremaine}(1999)}]{MT99}
{Magorrian}, J., \& {Tremaine}, S. 1999, \mnras, 309, 447

\bibitem[{{Mathews}(1989)}]{Mat89}
{Mathews}, W.~G. 1989, \aj, 97, 42

\bibitem[{{McKee \& Ostriker}(1977)}]{MO77}
{McKee}, C.~F., \& {Ostriker}, J.~P. 1977, \apj, 148

\bibitem[{{McNamara \& Nulsen}(2012)}]{MN12}
{McNamara}, B.~R., \& {Nulsen}, P.~E.~J. 2012, New Journal of Physics, 14, 5

\bibitem[{{Miller} {et~al.}(2012){Miller}, {Gallo}, {Treu}, \& {Woo}}]{MGT+12}
{Miller}, B., {Gallo}, E., {Treu}, T., \& {Woo}, J.-H. 2012, \apj, 747, 57

\bibitem[{{Muno et al.}(2004)}]{Mun+04}
{Muno}, M.~P., et al.\ 2004, \apj, 613, 326

\bibitem[{{Narayan \& Fabian}(2011)}]{NF11}
{Narayan}, R., \& {Fabian}, A.~C. 2011, \mnras, 415, 3721

\bibitem[{{Narayan et al.}(1998)}]{NMG+98}
{Narayan}, R., {Mahadevan}, R., {Grindlay}, J.~E., {Popham}, R.~G., \& {Gammie}, C. 1998, \apj, 492, 554

\bibitem[{{Narayan \& McClintock}(2008)}]{NM08}
{Narayan}, R., \& {McClintock}, J.~E. 2008, New Astronomy Reviews, 51, 733

\bibitem[{{Narayan \& Yi}(1994)}]{NY94}
{Narayan}, R., \& {Yi}, I. 1994, \apjl, 428, L13

\bibitem[{{Narayan et al.}(2000){Narayan}, {Igumenshchev}, \& {Abramowicz}}]{NIA00}
{Narayan}, R., {Igumenshchev}, I.~V., \& {Abramowicz}, M.~A. 2000, \apj, 539, 798

\bibitem[{{Pellegrini et al.}(2003)}]{PBF+03}
{Pellegrini}, S., {Baldi}, A., {Fabbiano}, G., \& {Kim}, D.-W. 2003, \apj, 597, 175

\bibitem[{{Pellegrini et al.}(2012)}]{PWF+12}
{Pellegrini}, S., {Wang}, J., {Fabbiano}, G., {Kim}, D.-W., {Brassington}, N.~J., {Gallagher}, J.~S., {Trinchieri}, G., \& {Zezas}, A. 2012, \apj, 758, 94

\bibitem[{{Proga \& Begelman}(2003){Proga}, \& {Begelman}}]{PB03}
{Proga}, D., \& Begelman, M.~C. 2003, \apj, 592, 767

\bibitem[{{Quataert}(2002)}]{Qua02}
{Quataert}, E. 2002, \apj, 575, 855

\bibitem[{{Quataert \& Gruzinov}(2000){Quataert}, \& {Gruzinov}}]{QG00}
{Quataert}, E., \& {Gruzinov}, A. 2000, \apj, 539, 809

\bibitem[{{Quataert \& Narayan}(2000)}]{QN00}
{Quataert}, E., \& {Narayan}, R. 2000, \apj, 528, 236

\bibitem[{{Randall et al.}(2006){Randall}, {Sarazin}, \& {Irwin}}]{RSI06}
{Randall}, S.~W., {Sarazin}, C.~L., \& {Irwin}, J.~A. 2006, \apj, 636, 200

\bibitem[{{Rees et al.}(1982){Rees}, {Begelman}, {Blandford}, \& {Phinney}}]{RBB+82}
{Rees}, M.~J., {Begelman}, M.~C., {Blandford}, R.~D., \& {Phinney}, E. S. 1982, \nat, 295, 17

\bibitem[{{Revnivtsev et al.}(2007){Revnivtsev}, {Churazov}, {Sazonov}, {Forman}, \& {Jones}}]{RCS+07}
{Revnivtsev}, M., {Churazov}, E., {Sazonov}, S., {Forman}, W., \& {Jones}, C. 2007, \aap, 473, 783

\bibitem[{{Revnivtsev et al.}(2008){Revnivtsev}, {Churazov}, {Sazonov}, {Forman}, \& {Jones}}]{RCS+08}
{Revnivtsev}, M., {Churazov}, E., {Sazonov}, S., {Forman}, W., \& {Jones}, C. 2008, \aap, 490, 37

\bibitem[{{Roberts et al.}(1991){Roberts}, {Hogg}, {Bregman}, {Forman}, \& {Jones}}]{RHB+91}
{Roberts}, M.~S., {Hogg}, D.~E., {Bregman}, J.~N., {Forman}, W.~R., \& {Jones}, C. 1991, \apjs, 75, 751

\bibitem[{{Sage \& Welch}(2006)}]{SW06}
{Sage}, L.~J., \& {Welch}, G.~A. 2006, \apj, 644. 850

\bibitem[{{Sazonov et al.}(2012)}]{SSR12}
{Sazonov}, S., {Sunyaev}, R., \& {Revnivtsev}, M. 2012, \mnras, 420, 388

\bibitem[{{Sharma et al.}(2011){Sharma}, {McCourt}, {Quataert}, \& {Parrish}}]{SMQ+11}
{Sharma}, P., {McCourt}, M., {Quataert}, E., \& {Parrish}, I.~J. 2011, \mnras, 420, 3174

\bibitem[{{Shcherbakov \& Baganoff}(2010)}]{SB10}
{Shcherbakov}, R.~V., \& {Baganoff}, F.~K. 2010, \apj, 716, 504

\bibitem[{{Shcherbakov et al.}(2013){Shcherbakov}, {Wong}, {Irwin}, \& {Reynolds}}]{SWI+13}
{Shcherbakov}, R.~V., {Wong}, K.-W., {Irwin}, J.~A., \& {Reynolds}, C.~S. 2013, \apj, submitted

\bibitem[{{Soker et al.}(2013){Soker}, {Akashi}, {Gilkis}, {Hillel}, {Papish}, {Refaelovich}, \& {Tsebrenko}}]{SAG+13}
{Soker}, N., {Akashi}, M., {Gilkis}, A., {Hillel}, S., {Papish}, O., {Refaelovich}, M., \& {Tsebrenko}, D. 2013, Astronomische Nachrichten, 334, 402

\bibitem[{{Su \& Irwin}(2013)}]{SI13}
{Su}, Y., \& {Irwin}, J. 2013, \apj, 766, 61

\bibitem[{{Syer \& Ulmer}(1999)}]{SU99}
{Syer}, D., \& {Ulmer}, A. 1999, \mnras, 306, 35

\bibitem[{{Tonry et al.}(2001) {Tonry}, {Dressler}, {Blakeslee}, {Ajhar}, {Fletcher}, {Luppino}, {Metzger}, \& {Moore}}]{TDB+01}
{Tonry}, J.~L., {Dressler}, A., {Blakeslee}, J.~P.,  {Ajhar}, E.~A., {Fletcher}, A.~B., {Luppino}, G.~A., {Metzger}, M.~R., \& {Moore}, C.~B. 2001, \apj, 546, 681

\bibitem[{{Tremblay et al.}(2012)}]{Tre+12}
{Tremblay}, G.~R. et al. 2012, \mnras, 424, 1026

\bibitem[{{Wilms et al.}(2000){Wilms}, {Allen}, \& {McCray}}]{WAM00}
{Wilms}, J., {Allen}, A., \& {McCray}, R. 2000, \apj, 542, 914

\bibitem[{{Wang \& Merritt}(2004)}]{WM04}
{Wang}, J., \& {Merritt}, D. 2004, \apj, 600, 149

\bibitem[{{Wang et~al.}(2012){Wang}, {Cheng}, \& {Li}}]{WCL12}
{Wang}, J.-M., {Cheng}, C., \& {Li}, Y.-R. 2012, \apj, 748, 147

\bibitem[{{Wang et~al.}(2013)}]{Wan+13}
{Wang}, Q.~D., et~al. 2013, Science, 341, 981

\bibitem[{{Wong et al.}(2011){Wong}, {Irwin}, {Yukita}, {Million}, {Mathews}, \& {Bregman}}]{WIY+11}
{Wong}, K.-W., {Irwin}, J.~A., {Yukita}, M., {Million}, E.~T., {Mathews}, W.~G., \& {Bregman}, J.~N. 2011, \apj, 736, L23 (W11)

\bibitem[{{Wong et al.}(2008){Wong}, {Sarazin}, {Blanton}, \& {Reiprich}}]{WSB+08}
{Wong}, K.-W., {Sarazin}, C.~L., {Blanton}, E.~L., \& {Reiprich}, T.~H.  2008 \apj, 682, 155

\bibitem[{{Wrobel \& Nyland}(2012)}]{WN12}
{Wrobel}, J.~M., \& {Nyland}, K. 2012, \aj, 144, 160

\bibitem[{{Yuan et~al.}(2002){Yuan}, {Markoff}, \& {Falcke}}]{YMF02}
{Yuan}, F., {Markoff}, S., \& {Falcke}, H. 2002, \aap, 383, 854

\bibitem[{{Yuan et~al.}(2003){Yuan}, {Quataert}, \& {Narayan}}]{YQN03}        
{Yuan}, F., {Quataert}, E., \& {Narayan}, R. 2003, \apj, 598, 301

\bibitem[{{Yuan et~al.}(2012){Yuan}, {Wu}, \& {Bu}}]{YWB12}
{Yuan}, F., {Wu}, M., \& {Bu}, D. 2012, \apj, 761, 129

\end{thebibliography}
\end{document}